\documentclass{elsarticle}
\usepackage[utf8]{inputenc}
\usepackage{amsmath}
\usepackage{amssymb}
\usepackage{float}
\usepackage{amsthm}
\usepackage{hyperref}
\usepackage{caption}
\usepackage{longtable}
\usepackage{array}
\usepackage{xcolor}
\usepackage{lscape}
\usepackage{tikz} 
\usepackage{subcaption}
\usepackage{afterpage}
\usepackage{rotating}
\usepackage[shortlabels]{enumitem}

\newcolumntype{C}[1]{>{\centering\arraybackslash}m{#1}}
\newtheorem{lem}{Lemma}
\newtheorem{thm}{Theorem}
\newtheorem{prop}{Proposition}
\newtheorem{defn}{Definition}
\newtheorem{remark}{Remark}

\usepackage{color}

\newcommand{\gal}{\text{Gal}}
\newcommand{\mypow}{\mbox{$a^{\bullet}$}}
\newcommand{\mysquare}{\mbox{\small{$\circledast$}}}
\newcommand{\calgo }{\kappa}
\newcommand{\ind}{\iota}

\newcommand{\revision}[1]{{{#1}}}
\begin{document}

\begin{frontmatter}

\title{New Results on Quasi-Subfield Polynomials}
 \author[1]{Marie Euler\corref{cor1}\fnref{fn1}}
 \ead{marie.euler@hotmail.com}
 \cortext[cor1]{Corresponding author}
 \fntext[fn1]{Part of this work was done in fulfillment of a master thesis requirement of the first author at the University of Oxford}
 \address[1]{DGA MI, France}

 \author[2,3]{Christophe Petit\fnref{fn2}\fnref{fn3}}
\ead{christophe.f.petit@gmail.com}
 \fntext[fn2]{Supported in part by  an EPSRC grant 	EP/S01361X/1.}
 \address[2]{Université libre de Bruxelles, Département d'informatique, Belgium}
 \address[3]{University of Birmingham, School of Computer Science, United Kingdom}

\date{May 2021}
\journal{Finite Fields and Their Applications}

\begin{abstract}
\footnote{Declarations of interest: none 

 © 2021. This manuscript version is made available under the CC-BY-NC-ND 4.0 license \url{http://creativecommons.org/licenses/by-nc-nd/4.0/}}
Quasi-subfield polynomials were introduced by Huang \emph{et~al.} together with a new algorithm to solve the Elliptic Curve Discrete Logarithm Problem (ECDLP) over finite fields of small characteristic.
In this paper we provide both new quasi-subfield polynomial families and a new theorem limiting their existence. Our results do not allow to derive any speedup for the new ECDLP  algorithm compared to previous approaches.
\end{abstract}
\end{frontmatter}

\section{Introduction}
Let $p$ be a prime and let $n,n'$ be two positive integers. For any prime power $q$, let $\mathbb{F}_{q}$ be the finite field with $q$ elements.  
When $n'$ divides $n$, the finite field $\mathbb{F}_{p^{n'}}$ is a subfield of $\mathbb{F}_{p^{n}}$. The polynomial
$$X^{p^{n'}}-X$$
splits over $\mathbb{F}_{p^{n}}$ and its roots are exactly all the elements of $\mathbb{F}_{p^{n'}}$.
Quasi-subfield polynomials, introduced by Huang \emph{et~al.}~\cite{quasi-subfield}, naturally generalize this polynomial.

\begin{defn}[informal]  \label{def:QSP1} A \emph{quasi-subfield polynomial (QSP)} is a polynomial of the form
$$L(X):=X^{p^{n'}}-\lambda(X)$$ such that ``most'' of its roots are distinct and defined over $\mathbb{F}_{p^{n}}$, and moreover $d:=\deg\lambda$ is ``small''. 
\end{defn}

When $n'$ does not divide $n$,  the degree of $\lambda$ cannot be too small, as shown in the following lemma.
\begin{lem}\emph{\cite[Lemma~4.1]{quasi-subfield}}\label{lem:MC}
Let $L(X)=X^{p^{n'}}-\lambda(X) \in \mathbb{F}_{p^n}[X]$ completely spliting over $\mathbb{F}_{p^{n}}$, such that $\ell:= \log_p \deg \lambda>0$. 
Then we have 
$$\left\lfloor \frac{n}{n'}\right\rfloor\ell + (n\bmod n') \geq  n'.$$
\end{lem}
\noindent A similar result in the non split case is also provided in~\cite[Lemma~C.2]{quasi-subfield}.

In light of this lemma, it is useful to associate to any QSP a ``quality'' parameter $\beta:=\ell n /n'^2$,
where $\ell:=\log_p\deg\lambda$ as above. \revision{This leads to a more formal definition of QSP :

\begin{defn}A \emph{quasi-subfield polynomial (QSP)} is a polynomial of the form  $$L(X):=X^{p^{n'}}-\lambda(X)\in\mathbb{F}_{p^n}[X]$$ which splits completely  (or at least has approximately $p^{n'}$ roots), for which moreover $\beta(L):=\dfrac{n\ell}{n'^2} \leq 1$ with $\ell:=\log_p{\deg\lambda}$.
\end{defn}}

In their paper, Huang \emph{et~al.} provide a QSP family over $\mathbb{F}_{p^{n}}[x]$ with $n=p_{a+1}$ and $\beta=1-\dfrac{1}{p_a}(1-1/p^r+1/(p^rp_a))$  with $p_a=1+p^r+p^{2r}+\cdots+p^{ar}$. Most importantly, they show how quasi-subfield polynomials can be used to solve the Elliptic Curve Discrete Logarithm Problem (ECDLP), a problem of major importance for cryptography.

\paragraph{Our results} We expand the study of quasi-subfield polynomials initiated in~\cite{quasi-subfield}, focusing on polynomials whose roots form a subgroup of either the additive or the multiplicative group of finite fields. 
In the additive case this amounts to searching for a \emph{linearized polynomial}
$$L(X)=X^{p^{n'}}-(a_{\ell}X^{p^{\ell}}+a_{\ell-1}X^{p^{\ell-1}}+\cdots+a_{0}X )$$
splitting over $\mathbb{F}_{p^n}$, with ``small'' $\ell$. In the multiplicative case this amounts to searching for a polynomial of the form
$$L(X)=X^{p^{n'}}-X^d$$
where $p^{n'}-d$ divides $p^n-1$, and $d$ is small.

Our main result is the following theorem on \emph{linearized} QSPs.
\begin{thm}
\label{thm:low_bound}
Let $L(X):=X^{p^{n'}}-(a_{\ell}X^{p^{\ell}}+a_{\ell-1}X^{p^{\ell-1}}+\cdots+a_{0}X )$ be a linearized polynomial over $\mathbb{F}_{p^n}$, with $\ell\geq1$. 
If $L$ splits completely  over $\mathbb{F}_{p^n}$  then $\beta:= \ell n /n'^2\geq 3/4$.

\end{thm}
\noindent 
The equality is obtained for example with $X^{p^{2}}+X^{p}+X$ in $\mathbb{F}_{p^{3}}[X]$.
The theorem improves on Lemma~\ref{lem:MC} for linearized polynomials with parameters $n,n'$ satisfying \revision{$(n\bmod n')\geq n'/4$. }
In the special case of trinomials, a similar result was independently obtained with similar techniques by McGuire and Mueller~\cite{mcguire2020some}.

We also introduce methods to generate new families of linearized QSP from known ones, and we exhibit new additive and multiplicative QSP families.

Finally, we apply our results to Huang \emph{et~al.}'s ECDLP algorithm.

\paragraph{Impact on ECDLP security} The complexity of Huang \emph{et~al.}'s ECDLP algorithm crucially relies on the quality parameter $\beta$ of the quasi-subfield polynomial used. Based on their complexity estimations, a value of $\beta$ at most 0.1 would be needed to obtain a complexity improvement over generic, state-of-the-art ECDLP algorithms.
In contrast, Theorem~\ref{thm:low_bound} rules out any $\beta$ smaller than 3/4 in the case of linearized polynomials, and the best quasi-subfield polynomial we found so far has $\beta \simeq 0.7$.
Further work will be needed to improve Huang \emph{et~al.}'s approach with new ideas and better QSP families, or to provide a definite proof that it will not improve on generic algorithms.

\paragraph{Outline} The remaining of this paper is organized as follows. Section~\ref{sec:bound} recalls our new lower bound for the $\beta$ value of linearized QSPs and it provides its proof. Section~\ref{sec:new_families} includes our new QSP families. Section~\ref{sec:applications} discusses the impact of our results on ECDLP and Section~\ref{sec:conclusion} concludes the paper.

\section{A new lower bound on $\beta$ for linearized quasi-subfield polynomials\label{sec:bound}}

In this section we first recall known properties of linearized polynomials, including a characterization of linearized polynomials that split completely over their field of definition. We then proceed to prove Theorem~\ref{thm:low_bound}, and we compare the bound it provides with the bounds given in~\cite{quasi-subfield} and~\cite{mcguire2020some}.

\subsection{Linearized polynomials}

Let $p$ be a prime and $n$ be positive integer.  Let $\mathbb{F}_{p^n}$ be the finite field with $p^n$ elements.
We write $\gal(\mathbb{F}_{p^n}/\mathbb{F}_p)$ for the Galois group of $\mathbb{F}_{p^n}$ with respect to $\mathbb{F}_p$.
For any automorphism $\sigma\in \gal(\mathbb{F}_{p^n}/\mathbb{F}_p)$, there exists $s\in\mathbb{Z}$ with  $\gcd(s,n)=1$ such that $\sigma(X)=X^{p^s}$. In the following we write $X^\sigma$ for $\sigma(X)$.

\begin{defn}[Linearized polynomials]
Let $\sigma\in \gal(\mathbb{F}_{p^{n}}/\mathbb{F}_p)$ and let $f=X^d+ a_{d-1}X^{d-1}+ \ldots +  a_1X+ a_0 \in \mathbb{F}_{p^n}[X]$.
The \emph{linearized polynomial} related to $f$ and $\sigma$ is the polynomial
$$L_{f,\sigma}=X^{\sigma^d}+  a_{d-1}x^{\sigma^{d-1}}+\ldots+a_1X^{\sigma} + a_0X  \in \mathbb{F}_{p^n}[X].$$ Moreover  $d$ is called the \emph{$\sigma$-degree} of $L_{f,\sigma}$. 
\end{defn}

Let $L:=L_{f,\sigma}
\in \mathbb{F}_{p^n}[X]$ be a linearized polynomial with coefficients in $\mathbb{F}_{p^n}$ as above. 
Let $C_L$ be the companion matrix of $f$, namely $$C_L:=\begin{bmatrix}
0 & 0 &\cdots &0& -a_0 \\
1 & 0 &\cdots & 0& -a_1 \\
0 &1 &\cdots & 0& -a_2 \\
\vdots & \vdots &\ddots & \vdots \\
0 & 0 &\cdots & 1 &-a_{d-1} \\
\end{bmatrix}.$$
We also define the matrix $$A_L:=C_L\cdot C_L^{\sigma}\cdot C_L^{\sigma^2}\cdots C_L^{\sigma^{n-1}},$$ where $C_L^{\sigma}$ is the matrix obtained by applying ${\sigma}$ coefficient-wise on $C_L$. Note that $A_L$ and $C_L$ are square matrices of dimension $d$. 

The following result (independently due to
McGuire and Sheekey~\cite{mcguire2019characterization} and Csajb{\'o}k \revision{\emph{et~al.}}~\cite{csajbok2019characterization}) characterizes linearized polynomials that split completely.

\begin{prop} \label{prop:splitMatrix}
Let $L=X^{\sigma^d}+  a_{d-1}x^{\sigma^{d-1}}+\ldots+a_1X^{\sigma} + a_0X$ be a linearized polynomial with coefficients in $\mathbb{F}_{p^n}$.
Then $L$ has $p^{n_1}$ roots defined over  $\mathbb{F}_{p^n}$, where $n_1$ is the dimension of the eigenspace of $A_L$ with eigenvalue $1$. In particular, $L$  splits completely over $\mathbb{F}_{p^n}[X]$ if and only if  $A_L$ is the identity matrix.
\end{prop}

A corollary of this proposition is that the maximum number of roots of $L$  is $p^d$, and $L_{f,\sigma}$ splits completely in $ \mathbb{F}_{p^n}$ only if $\sigma(X)=X^p$.
From now on, we will therefore only consider the case $\sigma(X)=X^p$ and thus write $L_f$ instead of $L_{f,\sigma}$.

Another important property of completely splitting linearized polynomials is the following one. 

\begin{prop}%[splitting Completely ]
\label{prop:comp_split}
Let $f=a_0 +a_1X +\cdots + X^d \in \mathbb{F}_{p}[X]$. Then the following properties are equivalent.
\begin{enumerate}
\item $L_{f}(X)$ splits completely over $\mathbb{F}_{p^n}[X]$
\item $L_{f}(X)$ divides $X^{p^n}- X$
\item $f$ divides $X^{n} - 1$ 
\end{enumerate}
\end{prop}
This proposition directly follows from the fact that $L_f$ divides $L_g$ for the composition of polynomials if and only if $f|g$ for the multiplication of polynomials~\cite[Chapter 11]{berlekamp1968algebraic}.

\subsection{Proof of Theorem~\ref{thm:low_bound} }
We will now prove Theorem~\ref{thm:low_bound}, namely that for any linearized polynomial $L=X^{p^{n'}}-(a_{\ell}X^{p^{\ell}}+a_{\ell-1}X^{p^{\ell-1}}+\cdots+a_{0}X )$ with $\ell\geq 1,~a_{\ell}\neq 0$ and $\forall i, a_i \in \mathbb{F}_{p^n}$, if $L$ splits  completely over $\mathbb{F}_{p^n}$  then $\beta := \ell n /n'^2\geq 3/4$.

This result is in fact a consequence of the following lemma which highlights that the field has to be big enough to have completely splitting sparse linearized polynomials in it.

\begin{lem}[Lower bound on $n$]
\label{lem:low_bound}
Let $L_f=X^{p^{n'}}-(a_{\ell}X^{p^{\ell}}+a_{\ell-1}X^{p^{\ell-1}}+\cdots+a_{0}X)$ with $\ell\geq 1, ~a_{\ell}\neq 0$ and $\forall i, a_i \in \mathbb{F}_{p^n}$. If $L\revision{_f}$ splits completely over $\mathbb{F}_{p^n}$  then 
$$ n \geq n' + (n'-\ell)\Bigl\lfloor \dfrac{n'-1}{\ell}\Bigr\rfloor $$
\end{lem}

Let us first observe that this result is indeed enough to prove the theorem.
Indeed one can  notice that since $n'$ and $\ell$ are integers, we have
$\Bigl\lfloor \dfrac{n'-1}{\ell}\Bigr\rfloor \geq \dfrac{n'}{\ell}-1$.
Therefore, by Lemma~\ref{lem:low_bound} we have $n \geq n'+ (n'-\ell) \dfrac{n'}{\ell}-(n'-\ell) \geq \dfrac{n'^2}{\ell}-n'+\ell$. 
Thus, $\beta = \dfrac{\ell n}{n'^2}\geq 1- \dfrac{\ell}{n'}+\dfrac{\ell^2}{n'^2} = 1 - \dfrac{\ell}{n'}\left(1-\dfrac{\ell}{n'}\right)\geq 1-1/4= 3/4$.

We will now show that Lemma~\ref{lem:low_bound} boils down to the proof of a result about the power of a matrix. By Proposition~\ref{prop:splitMatrix}, $\revision{L:=L_f}$ splits  completely  over $\mathbb{F}_{p^n}$  if and only if  $A_L:=C_L\cdot C_L^{\sigma}\cdots C_L^{\sigma^{n-1}}=I$ where $C_L$ is the companion matrix of $f$.

Therefore we have to prove that $$  \text{If } n< n' + (n'-\ell)\Bigl\lfloor \dfrac{n'-1}{\ell}\Bigr\rfloor, \text{ then } \  A_{L}\neq I$$

The remainder of the proof will consider matrices defined over the polynomial ring $\mathbb{F}_{p^n}[a_0,\ldots,a_\ell]$, where $a_i$ are the coefficients of $L$ as above. 
However as our result only depends on the value of $n'$ and $\ell$ but not on the specific coefficients $a_i$, \revision{we represent  any element $x\in\mathbb{F}_{p^n}[a_i]$ by a symbol $\tilde x \in  \left\{\mathit{0}, \mathit{1}, \mypow, \mysquare \right\}$, noting it $x \leadsto \tilde x$, according to the following rules :
\begin{itemize}
    \item if $x \leadsto 0$ then $x=0$;
    \item if $x \leadsto 1$ then $x=1$;
    \item if $x \leadsto \mypow$ then $x$ is a power of $(-a_{\ell})$ and thus $x\neq 0$;
    \item ($x \leadsto \mysquare$ implies no condition on $x$).
\end{itemize}

For instance $-a_0 \leadsto \mysquare$, $- a_{\ell} \leadsto \mypow $, $a_0 + a_{\ell} \leadsto \mysquare$, and also $-a_{\ell} \leadsto \mysquare$. One can notice that if $x_1\leadsto \tilde{x_1}$ and $x_2 \leadsto \tilde{x_2}$, then $x_1+ x_2 \leadsto z_1$ and $x_1 \cdot x_2 \leadsto z_2$ with $z_1$ and $z_2$ given by the following tables.}

$$\begin{array}{|c|cccc|c|c|cccc|}

\cline{1-5} \cline{7-11}
+    & 0 & 1 & \mypow & \mysquare &  & \cdot & 0 &  1 & \mypow & \mysquare \\ 	\cline{1-5} \cline{7-11}
0    & 0 & 1 & \mypow & \mysquare &  &   0    &0 & 0 &   0    & 0 \\
1    & 1 & \mysquare & \mysquare & \mysquare &  &1 &  0 &  1 & \mypow & \mysquare  \\
\mypow & \mypow &\mysquare & \mysquare & \mysquare & & \mypow & 0 & \mypow &  \mypow & \mysquare \\
\mysquare    & \mysquare &  \mysquare & \mysquare    & \mysquare &  &   \mysquare    & 0 &  \mysquare&  \mysquare    & \mysquare \\ 	\cline{1-4} \cline{1-5} \cline{7-11}

\end{array}$$

\revision{For instance, if $x_1 \leadsto \mypow$, $x_2 \leadsto \mypow$, $x_3 \leadsto \mysquare$, then $x_1 + x_2 \leadsto \mysquare$, $x_1\cdot x_2 \leadsto \mypow$ and $x_1\cdot x_3 \leadsto \mysquare$.}

 \revision{We extend this notation to matrices over $\mathbb{F}_{p^n}[a_i]$: Let $M=(m_{i,j})$ be a square matrix of dimension $n'$ with coefficients in $\mathbb{F}_{p^n}$. If for all $i$ and $j$, $m_{i,j} \leadsto \tilde{m}_{i,j}$,  we denote $M \leadsto \tilde{M} := (\tilde{m}_{i,j})$. We observe that if $M\leadsto \tilde{M}$ and $N\leadsto \tilde{N}$ then $M+N \leadsto \tilde{M}+\tilde{N}$ and $MN \leadsto \tilde{M}\tilde{N}$ where the operations are done according to the above tables.} 

Let $M$ and $\tilde{M}$ such that $M\leadsto \tilde{M}$. 
If there is a zero on the main diagonal of $\tilde{M}$ or there is a non zero coefficient (1 or $\mypow$) outside of this diagonal, then we see that $\tilde{M}$ cannot represent the identity matrix (denoted $I\not \leadsto \tilde{M}$) and  then $M\neq I$.

Therefore we will give  $\tilde{A_L}$ such that $A_L \leadsto \tilde{A_L}$ and prove that if $n < n' + (n'-\ell)\Bigl\lfloor \dfrac{n'-1}{\ell}\Bigr\rfloor$ then $I\not\leadsto \tilde{A_L}$.

First of all, we can recall that $A_L:=C_L\cdot C_L^{\sigma}\cdot\cdots\cdot C_L^{\sigma^{n-1}}$ with
 {\small$$C_L=\begin{bmatrix}
0 & 0 &\cdots & 0& & \cdots & \cdots& -a_0 \\
1 & 0 &\cdots & 0& &\cdots& \cdots & -a_1 \\
0 &1 &\cdots & 0& &\cdots & \cdots & -a_2 \\
\vdots & \vdots &\ddots & & &\ddots  & &\vdots \\
0 & 0 &\cdots & 1 & & \cdots  &\cdots & -a_{\ell} \\
\vdots & \vdots &&  \vdots &\ddots & & &0 \\
\vdots & \vdots &  & \vdots & &\ddots  && \vdots \\
0 & 0 &\cdots & & \cdots &  &  1 &0 \\
\end{bmatrix}$$} 

Moreover, since $\sigma$ acts on matrices coefficient-wise, we can observe that for all $k\geq 0$,  $C_{L}^{\sigma^{k}}\leadsto M$ with 

$$M:=\begin{bmatrix}
0 & 0 &\cdots & 0& & \cdots & \cdots& \mysquare \\
1 & 0 &\cdots & 0& &\cdots& \cdots & \mysquare \\
0 &1 &\cdots & 0& &\cdots & \cdots & \mysquare \\
\vdots & \vdots &\ddots & & &\ddots  & &\vdots \\

0 & 0 &\cdots & 1 & & \cdots  &\cdots & \mypow \\
\vdots & \vdots &&  \vdots &\ddots & & &0 \\
\vdots & \vdots &  & \vdots & &\ddots  && \vdots \\
0 & 0 &\cdots & & \cdots &  &  1 &0 \\
\end{bmatrix} $$

Therefore $A_{L} \leadsto M^n$. Our goal is then to study the powers of $M$, which is a companion matrix defined on $\left\{0,1,\mypow,\mysquare\right\}$ and to prove the following result.

\begin{lem}[Small powers are not the identity]
\label{lem:powers_M}
$$\text{If }  n < n' + (n'-\ell)\Bigl\lfloor \dfrac{n'-1}{\ell}\Bigr\rfloor,\ \text{ then }  I\not \leadsto M^{n}.$$
\end{lem}

\noindent More precisely, we will prove the following lemma which exhibits a non zero coefficients outside of the main diagonal. We will note $M^n_{i,j}$ the coefficient in the $i$-th row, $j$-th column of the matrix $M^n$.

\begin{lem}[A non zero coefficient] 
\label{lem:non_null_coeff}
The following two claims are true:

\begin{itemize}
\item If $n< n'$, then $M^{n}_{n+1,1}=1 $.
\item If  $n'\leq n < n' + (n'-\ell)\Bigl\lfloor \dfrac{n'-1}{\ell}\Bigr\rfloor$, then we have $M^{n}_{i_{n},1}=\mypow\neq 0 $ \\
with {$ i_{n}= n-(n'-\ell)\Bigl\lfloor \dfrac{n -\ell}{n'-\ell}\Bigr\rfloor+1 \in \left[2,n'\right]$}  \end{itemize}

\end{lem}

\begin{proof}
Adapting a result of Chen and Louck~\cite{chen_combinatorial_1996} \revision{(see~ \ref{app:powers_of_companion} for details)} on powers of companion matrices we have $M^{n}_{i,j}=1$ if $n=i-j$
and otherwise
$$M^{n}_{i,j}=\sum\limits_{\substack{\textbf{k}=(k_1,\cdots,k_{n'})\\ \sum\limits_{\ind=1}^{n'}\ind k_{\ind}=n -i+j}}w_{\textbf{k}}\cdot
0^{k_1+\cdots+k_{n'-\ell-1}} \cdot (\mypow)^{k_{n'-\ell}}\cdot \mysquare^{k_{n'-\ell+1}+\cdots+k_{n'}} $$
where $\textbf{k}=(k_{\ind})_{1\leq \ind\leq n'}$ are non-negative integers and $$w_{\textbf{k}}=\dfrac{k_{n'-i+1}+\cdots+k_{n'}}{k_{1}+\cdots+k_{n'}}\binom{k_{1}+\cdots+k_{n'}}{k_{1},\cdots,k_{n'}}.$$
The first part of Lemma~\ref{lem:non_null_coeff} follows from the case $n=i-j$.

When $n\neq i-j$, the contribution of any \textbf{k} with $k_1+\cdots+k_{n'-\ell-1}>0$ to the sum  is null. Therefore  we get 
\begin{equation}\label{eq:1}
M^{n}_{i,j}=\sum\limits_{\substack{\textbf{k}=(k_{n'-\ell},\cdots,k_{n'})\\\sum\limits_{\ind=n'-\ell}^{n'}\ind k_{\ind}=n -i+j}}w_{\textbf{k}}\cdot \mypow \cdot \mysquare^{k_{n'-\ell+1}+\cdots+k_{n'}},
\end{equation}
where $w_{\textbf{k}}$ is now 
$$w_{\textbf{k}}=\dfrac{k_{\max(n'-i+1,n'-\ell)}+\cdots+k_{n'}}{k_{n'-\ell}+\cdots+k_{n'}}\binom{k_{n'-\ell}+\cdots+k_{n'}}{k_{n'-\ell},\cdots,k_{n'}}.$$

Note that here we do not want to remove the exponents over the unknown symbol. Indeed when $k_{n'-\ell+1}+\cdots+k_{n'}=0$, we have $\mysquare^{k_{n'-\ell+1}+\cdots+k_{n'}}=1$; so we get a term $w_{\textbf{k}}\cdot \mypow $ which in the case  $w_{\textbf{k}}=1$ is exactly what we need to prove that this coefficient is \mypow.

Let $n'\leq n \leq n'+(n'-\ell)\Bigl\lfloor \dfrac{n'-1}{\ell} \Bigr\rfloor -1$, and let $i_{n}:=	n-(n'-\ell)\Bigl\lfloor \dfrac{n -\ell}{n'-\ell}\Bigr\rfloor+1$. We have
\begin{align*}
	n-(n'-\ell) \dfrac{n -\ell}{n'-\ell} +1& \leq i_{n}   \leq 	n-(n'-\ell) \left(\dfrac{n -\ell}{n'-\ell}- \dfrac{n' -\ell-1}{n'-\ell}\right) +1 \\
\revision{n-(n-\ell)+1} & \leq i_n  \leq n -(n -n'+1)  +1\\
\revision{2\leq \ell+1} & \leq i_n  \leq n'\\
\end{align*}

In particular, $M^{n}_{i_{n},1}$ is well-defined and it is not in the top left corner.

Moreover, $n'-	i_{n}+1\leq n'-(\ell+1)+1=n'-\ell$ so $\max(n'-	i_{n}+1,n'-\ell)=n'-\ell$. Also,  $n-i_n+1=(n'-\ell)\Bigl\lfloor \dfrac{n -\ell}{n'-\ell}\Bigr\rfloor$.
We then have :
\begin{align}
	w_{\textbf{k}}= \frac{k_{n'-\ell}+\cdots+k_{n'}}{k_{n'-\ell}+\cdots+k_{n'}}\binom{k_{n'-\ell}+\cdots+k_{n'}}{k_{n'-\ell},\cdots,k_{n'}}&=&\binom{k_{n'-\ell}+\cdots+k_{n'}}{k_{n'-\ell},\cdots,k_{n'}} \nonumber\\ 
	& =& \dfrac{(k_{n'-\ell}+\cdots+k_{n'})!}{k_{n'-\ell}!\cdots k_{n'}!}  \nonumber\end{align}
	and 
	\begin{equation*}
	M^{n}_{i_{n},1}   
=\sum_{\substack{k_{n'-\ell},\cdots,k_{n'}\\\sum\limits_{\ind=n'-\ell}^{n'}\ind k_{\ind}=(n'-\ell)\Bigl\lfloor \dfrac{n -\ell}{n'-\ell}\Bigr\rfloor}} w_{\textbf{k}} \cdot \mypow\cdot \mysquare^{k_{n'-\ell+1}+\cdots+k_{n'}}
\end{equation*}

In order to show that $	M^{n}_{i_{n},1}=\mypow$, we now show that this sum in fact only involves one term, and that this term is exactly \mypow. 

\revision{
Clearly $\textbf{k}=(k_{n'-\ell},\cdots,k_{n'})=\left(\Bigl\lfloor \dfrac{n -\ell}{n'-\ell}\Bigr\rfloor ,0,\cdots,0\right)$ is a solution to 
\begin{equation} \label{eq:eq2}
    \sum\limits_{\ind=n'-\ell}^{n'}\ind k_{\ind}=(n'-\ell)\Bigl\lfloor\dfrac{n -\ell}{n'-\ell}\Bigr\rfloor 
\end{equation}
and for this $\textbf{k}$ we have $w_{\textbf{k}} \cdot \mypow\cdot \mysquare^{k_{n'-\ell+1}+\cdots+k_{n'}}=1\cdot\mypow\cdot 1 =\mypow$. It remains to show that it is the only valid $\textbf{k}$ to prove that $M^{n}_{i_{n},1}=\mypow$.

By contradiction, let us assume that there exists a solution $(k_{n'-\ell},\cdots,k_{n'})$ to Equation~(\ref{eq:eq2}) such that
\begin{align}
\sum\limits_{\ind=n'-\ell+1}^{n'}k_{\ind}&=\sum\limits_{\ind=0}^{\ell-1}k_{n'-\ind}>0 \label{eq:hyp1}
\end{align}}
 We recall that by definition, all the $k_{\ind}$ are non-negative.
From \revision{Equation (\ref{eq:eq2})}, we have :
\begin{align}
 (n'-\ell)\Bigl\lfloor \dfrac{n -\ell}{n'-\ell}\Bigr\rfloor&=\sum\limits_{\ind=n'-\ell}^{n'} \ind  k_{\ind}=\sum\limits_{\ind=0}^{\ell}(n'-\ind)k_{n'-{\ind}}\nonumber\\ 
 &=(n'-\ell)\sum_{\ind=0}^{\ell}k_{n'-\ind}+ \sum_{\ind=0}^{\ell-1}(\ell-\ind)k_{n'-\ind}\label{eq:eq3}\\
&\geq (n'-\ell)\dfrac{\sum_{\ind=0}^{\ell}(\ell-\ind)k_{n'-\ind}}{\ell}+ \sum_{\ind=0}^{\ell-1}(\ell-\ind)k_{n'-\ind}\nonumber \\
&\geq \left( \dfrac{n'-\ell}{\ell}+ 1\right) \sum_{\ind=0}^{\ell-1}(\ell-\ind)k_{n'-\ind} = \dfrac{n'}{\ell} \sum_{\ind=0}^{\ell-1}(\ell-\ind)k_{n'-\ind}\label{eq:eq4}
\end{align}

 From Equation~\eqref{eq:eq3},  we have $(n'-\ell)|\sum_{\ind=0}^{\ell-1} (\ell-\ind)k_{n'-\ind}$.
We also have, by \revision{hypothesis~\eqref{eq:hyp1}}, that $\sum_{\ind=0}^{\ell-1} (\ell-\ind)k_{n'-\ind}\geq \sum_{\ind=0}^{\ell-1} k_{n'-\ind}>0$ hence $\sum_{\ind=0}^{\ell-1} (\ell-\ind)k_{n'-\ind}\geq (n'-\ell)$.

 Together with Equation~\eqref{eq:eq4} this implies
$(n'-\ell)\Bigl\lfloor \dfrac{n -\ell}{n'-\ell}\Bigr\rfloor \geq  \dfrac{n'}{\ell} (n'-\ell)$ so
\begin{equation}\label{eq:contradiction1}
\Bigl\lfloor \dfrac{n -\ell}{n'-\ell}\Bigr\rfloor \geq 
\dfrac{n'}{\ell}
.\end{equation}

On the other hand, thanks to the condition on $n$ being not too big, we have 
$$\dfrac{n-\ell}{n'-\ell}\leq \dfrac{n'+(n'-\ell)\Bigl\lfloor \dfrac{n'-1}{\ell} \Bigr\rfloor -1 -\ell }{n'-\ell} \leq \Bigl\lfloor \dfrac{n'-1}{\ell}\Bigr\rfloor +\left(1-\dfrac{1}{n'-\ell}\right).$$

This implies $\Bigl\lfloor\dfrac{n-\ell}{n'-\ell} \Bigr\rfloor \leq  \Bigl\lfloor \dfrac{n'-1}{\ell} \Bigr\rfloor < \dfrac{n'}{\ell}$, contradicting Equation~\eqref{eq:contradiction1}. We deduce that $\textbf{k}=(\Bigl\lfloor \dfrac{n -\ell}{n'-\ell}\Bigr\rfloor ,0,\cdots,0)$ is the only solution to Equation~\eqref{eq:eq2} and thus $$M^{n}_{i_{n},1}=\mypow\neq 0$$
\end{proof}

\subsection{Comparison with other lower bounds}
Theorem~\ref{thm:low_bound} improves on the bound given in Lemma~4.1 of \cite{quasi-subfield}  whenever the QSP is linearized and $(n\bmod n')\geq n'/4$.

Our bound is similar to the one given by Daniela Mueller and Gary McGuire in \cite{mcguire2020some}, which was established during the completion of this article. Indeed, Theorem 1.1 in~\cite{mcguire2020some} shows (using also \cite{mcguire2019characterization} and \cite{csajbok2019characterization}) that a linearized trinomial $L=X^{q^{d}}-b X^q-aX \in \mathbb{F}_{p^n}$, with $b\neq0$ and $q=p^k$ a power of $p$ such that $n=k\tilde{n}$,
splits completely only if 
$$\tilde{n} \geq (d-1)d +1 =d^2-d+1.$$

Let us compare it with the bound given by  Lemma~\ref{lem:low_bound}.
We write $n=k\tilde{n}$ so that $q=p^k$. Thus $L=X^{p^{kd}}-b X^{p^k}-aX $. Lemma~\ref{lem:low_bound} gives: 
$$k\tilde{n}=n\geq kd +(kd-k) \Bigl\lfloor \dfrac{kd-1}{k}\Bigr\rfloor \geq k(d+(d-1)(d-1))$$
\noindent Thus, as Mueller and McGuire, we get $\tilde{n}\geq d+(d-1)^2=d^2-d+1 $.

While their bound is less general than ours as it only applies to trinomials, they gave a more complete description of completely splitting linearized trinomials. \revision{Indeed, they also take into account the case $\ell=0$ which we did not consider in this article, and they exhaustively describe all possible such polynomials when $\tilde{n} \leq (d-1)d +1$:
\begin{itemize}
\item either $\tilde{n}=id$ with $i\leq d-1$, $b=0$ and $a^{1+q^d+\cdots+q^{(i-1)d}}=1$,
\item either $\tilde{n}= (d-1)d +1$, $a^{1+q+\cdots+q^{(d-1)d}}=(-1)^{d-1}$, $b=-a^{qe_1}$ where $e_1=\sum_{i=0}^{d-1}q^{id}$ and  $d-1$ is a power of $p$
\end{itemize}
 
See also \cite{santonastaso2020linearized} for the complete description of completely splitting linearized trinomials when $\tilde{n} \leq (d-1)d +d-1$.}

\section{New families of quasi-subfield polynomials \label{sec:new_families}}

In this section, we will prove that the additive polynomials of Proposition~\ref{prop:add_families} and the multiplicative polynomials of Proposition~\ref{prop:mult_families} are quasi-subfield polynomials. 
We first provide general tools to deduce new linearized QSPs from known ones and we define equivalence classes among  linearized quasi-subfield polynomials. We then successively focus on additive and multiplicative families, and we finish with the case where the extension degree is a Mersenne prime.

Since the case $\log_p \deg \lambda =0$ corresponds to subfield polynomials, which are well-known, we will only consider the case $\log_p \deg \lambda >0$.

\begin{table}
{\small
\renewcommand{\arraystretch}{2}
\begin{tabular}{|C{3.8cm}|C{2cm}|C{1.25cm}|C{3.5cm}|}
\hline
Quasi-subfield polynomial & $p$& $n$ & $\beta$ \\
\hline
\hline
$L_{f_a}$ with $f_0= 1+X^{q-1}+\cdots+X^{q^d-1} $, $f_a=a+X+X^q+\cdots +X^{q^d}$ for any $a\neq0\in \mathbb{F}_q$ and $q$ a power of $p$  & any & $q^{d+1}-1$ &$ 1 - \frac{q^{d-1}}{(1+q+\cdots+q^{d-1})^2} $ \\

\hline
$L_{(X^n-1)/f}$ with $L_f$ a completely splitting QSP of degree n' & any & any integer & \revision{$ 1- (\frac{n'}{n-n'})^2(1-\beta(L_f))$}\\

\hline
$X^{p^{n'}}-X^a$ with $n'=n-i$  and {$a=p^{n'} \mod (\frac{p^{n}-1}{p^{2i}-1})$ }& any & $2ik$, $k\geq 2$ & $\leq 1-\dfrac{1}{(2k-1)^2}$ \\

\hline
$X^{p}-X^a$ with $k\geq 2 $  and {$a=p \mod (\dfrac{p-k}{k-1})$}& $k^n+k-1$  & any integer & $\leq 1$ \\

\hline
$X^{p^{n-1}}-X^a$ with $k\geq 2$ and $r=\frac{(p^n-1)(k-(-1)^n)}{(k^{n}-k)(k^n-(-1)^n)}$ $a=p^{n-1} \mod r$ & $k^n-k-(-1)^n$ & $n>2$  $k^n \gg1$ &$ \leq 1 $\\

\hline
\end{tabular}
}\caption{New families of quasi-subfield polynomials\label{tab:qsp}}
\end{table}

\subsection{Equivalent classes of linearized quasi-subfield polynomials}
In order to simplify our search of quasi-subfield polynomials, we will first discuss transformations to deduce new  linearized quasi-subfield polynomials from known ones. We will introduce two types of transformations. The first transformation will change the value of $\beta$ and thus potentially improve it. The second one will keep the same $\beta$, thus it will not produce more interesting linearized quasi-subfield polynomials, but it will allow us to group them by equivalence classes. 

Both transformations will only concern completely splitting QSPs \revision{with coefficients in $\mathbb{F}_p$}.
We will write \revision{$\mathcal{Q}^L_{p,n}$} for the set of  completely splitting linearized QSPs  in $\mathbb{F}_{p^n}[X]$.

The first way to obtain a linearized quasi-subfield polynomial from another one is what we call the inversion process.

\begin{prop}[Inversion]
\label{prop:inversion}
Let $f=X^{n'}+a_{\ell}X^{\ell}+\cdots+a_0\in \mathbb{F}_{p}[X]$ 
such that $L_{f} \in \mathcal{Q}^L_{p,n}$ and $n'<n$. Let $g=(X^n-1)/f$. Then $L_g \in \mathcal{Q}^L_{p,n}$. We say that $L_g$ is the inverse of $L_f$.
\end{prop}

\begin{proof}
By Proposition~\ref{prop:comp_split}, we know that $f|X^n-1$ so $g$ is well-defined.

\revision{Let us write $g$ as $X^{n-n'}+b_{r}X^{r}+\cdots+b_0$. We will prove that $r+n' = n-n'+\ell$. Indeed, $X^n-1=f\cdot g= (X^{n'}+a_{\ell}X^{\ell}+\cdots+a_0)(X^{n-n'}+b_{r}X^{r}+\cdots+b_0)=X^n+b_rX^{r+n'}+a_{\ell}X^{\ell+n-n'}+ \cdots+ a_0b_0$. So, if $r+n'>\ell+n-n'$ then the coefficient of $X^{r+n'}$ in $f\cdot g$ comes exclusively from $(b_r X^r)\cdot X^{n'}$ and is therefore non zero. Conversely, if $r+n'<\ell+n-n'$ then the coefficient of $X^{\ell+n-n'}$ in $f\cdot g$ comes exclusively from $(a_{\ell} X^{\ell})\cdot X^{n-n'}$ and thus is not zero. 
We deduce
\begin{align*}
\beta(L_g)=\frac{n.r}{(n-n')^2}   & = \frac{n.(n-2n'+\ell)}{(n-n')^2}=1-\frac{n'^2-\ell.n}{(n-n')^2} \\
&= 1-\left(\frac{n'}{n-n'}\right)^2(1-\beta(L_f))\leq 1.
\end{align*} }

\end{proof}

Our second family of transformations keep the value of $\beta$ unchanged.

\begin{prop}[Transformations preserving $\beta$]

\label{prop:k_property}

Let $k\geq 1$ and $\gamma \in \mathbb{F}_{p^n}^*$.  Let $f=X^{n'}+a_{\ell}X^{\ell}+\cdots+a_0\in \revision{\mathbb{F}_{p}}[X]$. 
Then the following properties are equivalent:
\begin{enumerate}[(a)]
\item $L_{f} \in \mathcal{Q}^L_{p,n}$,
\item $L_{f(X^k)}=X^{p^{k.n'}}+a_{\ell}X^{p^{k.\ell}}+\cdots+a_1X^{p^k}+a_0X \in \mathcal{Q}^L_{p,kn}$,
\item \revision{(when $n|p-1$) for any $\alpha \in \mathbb{F}_{p}$ with $\alpha^n=1$, }$\alpha^{-n'} L_{f(\alpha.X)}= \alpha^{-n'}(\alpha^{n'}X^{p^{n'}}+\alpha^{\ell}a_{\ell}X^{p^{\ell}}+\cdots+\alpha a_1X^p+a_0X) \in \mathcal{Q}^L_{p,n}$,
\item $\gamma^{-p^{n'}}L_{f}(\gamma.X)=\gamma^{-p^{n'}}((\gamma.X)^{p^{n'}}+a_{\ell}(\gamma.X)^{p^{\ell}}+\cdots+a_1(\gamma.X)^{p}+a_0\gamma X )\in\mathcal{Q}^L_{p,n}$.
\end{enumerate}
\end{prop}

\begin{proof} 
One can observe that these four quasi-subfield polynomials have the same $\beta$. Therefore, we only have to show that their splitting conditions are equivalent.
The equivalence between (a) and (b) directly comes  from Proposition~\ref{prop:comp_split}. Indeed,
\begin{align*}
 L_{f}\in \mathcal{Q}^L_{p,n} \Leftrightarrow  f|X^n-1 & \Leftrightarrow  f(X^k)|X^{kn}-1  \text{ in }  \mathbb{F}_{p^n}[X]\\
& \Leftrightarrow L_{f(X^k)} \in \mathcal{Q}^L_{p,kn}.
\end{align*}
Properties (a) and (c) are also equivalent as 
\begin{align*}
 L_{f}\in \mathcal{Q}^L_{p,n} \Leftrightarrow  f|X^{n}-1 & \Leftrightarrow  f(\alpha.X)|(\alpha.X)^{n}-1 \\
& \Leftrightarrow  f(\alpha.X)| X^n-1 \text{ since } \alpha^n=1\\
& \Leftrightarrow  \alpha^{-n'}f(\alpha.X)| X^n-1 \\
& \Leftrightarrow  \alpha^{-n'}L_{f(\alpha.X)} \in \mathcal{Q}^L_{p,n}. 
\end{align*}
Finally, replacing $X$ by $\gamma\cdot X$ clearly does not change the fact that the polynomial is split, therefore $(a)\Leftrightarrow (d)$ is trivial.
\end{proof}

\revision{Let us now reconsider the transformations $(a) \Leftrightarrow (b) $ and $(a) \Leftrightarrow (c)$. As these transformations do not change the value of $\beta$ and they send completely splitting linearized quasi-subfield polynomials with coefficients in $\mathbb{F}_p$ onto other ones, we can define equivalence classes by saying that two  completely splitting linearized quasi-subfield polynomials with coefficients in $\mathbb{F}_p$} are equivalent to each other if one can be obtained from the other with one through one of the previous transformations. Obviously, since the transformations leave the value of $\beta$ - which determines the efficiency of the ECDLP algorithm - unchanged,  we are only interested in finding one representative of each class.

\subsection{Examples of completely splitting linearized QSPs}
\label{sec:systematic_search}
In order to find examples of linearized QSPs, we performed a systematic search of representatives of classes of equivalence of completely splitting QSPs. We will now explain how we did this search, and present our results.

 From now on, we will only consider polynomials with coefficients in the base field $\mathbb{F}_p$. 
Then any $L$ of the shape $X^{p^{kn'}}+a_{\ell}X^{p^{k\ell}}+\cdots+a_0\in \mathcal{Q}^L_{p,kn}$ is equivalent (in the above sense) to  $X^{p^{n'}}+a_{\ell}X^{p^{\ell}}+\cdots+a_0\in \mathcal{Q}^L_{p,n}$. Whenever we have a non trivial factor $d$ of all the element of the set $\left\{i \geq 1, a_i \neq 0 \right\}\cup\left\{n\right\}$, we may use it to reduce the degree of the polynomial by a factor d. Therefore we may reduce the search of representatives of each class to polynomials of the shape $X^{p^{n'}}+a_{\ell}X^{p^{\ell}}+\cdots+a_0 \in \mathbb{F}_{p^n}[X] $ with $\left\{i \geq 1, a_i \neq 0 \right\}\cup\left\{n\right\}$ setwise coprime. 

One can also notice that transformation $(a) \Leftrightarrow (c)$ cannot often be used. Indeed if $n$ is prime then $\alpha^n=1$ implies $n|p-1$.

Since we restrict the search to polynomials in $\mathbb{F}_p[X]$, all the coefficients of $C_L$ are in $\mathbb{F}_p$ and thus $A_L$ is merely $C_L^n$. Hence, Proposition~\ref{prop:splitMatrix} says that $L$ splits completely  over $\mathbb{F}_{p^n}$ if and only if $C_L^n=I$.

Let $L=X^{p^{n'}}+a_{\ell}X^{p^{\ell}}+\cdots+a_0$ be a fixed linearized polynomial in $\mathbb{F}_p[X]$. We can assume $a_0\neq 0$ since if $a_0=0$ then $0$ is a root of $L$ with multiplicity at least $p$ so $L$ does not split completely. We can search for the smallest $n$ such that $L$ splits completely  over $\mathbb{F}_{p^n}$. This amounts to searching for $n$ such that $C_L^n=I$, in other words finding the order of $C_L$. Note that $C_L$ is in $\text{GL}_{n'}(\mathbb{F}_p)$  since $\det{C_L}=(-1)^{n'}a_0\neq 0$, hence $n$ exists.  Moreover as we also want $\beta(L)=n.\ell/(n')^2\leq 1$, we only have to compute $C_L^k$ with $k<n'^2/\ell$. If we find such a $k$ with $C_L^k=I$ then $L_f\in \mathcal{Q}^{L}_{p,k}$.

This naturally leads to an algorithm to produce a set of representatives of the previously defined equivalence classes. 
\revision{The results output by our algorithm when asking for representatives of the equivalence classes for $p\in\{2,3,5,7\}$, $n'\leq16$ and coefficients values restricted to $\left\{0,1,-1\right\}$, are presented in appendix (Table~\ref{tab:linear_qsp}). Observing patterns in them allowed us to conjecture new types of quasi-subfield polynomials. We present one representative per equivalence class, as other quasi-subfield polynomials can be obtained by using the rules listed in Proposition~\ref{prop:k_property}.}

\begin{prop}[Families of linearized QSPs]
\label{prop:add_families}
The following types of linearized polynomials are quasi-subfield polynomials:
\begin{description}
\item[Type 1] $L_h$ with $h=X^{p_a}+\cdots+X^{p_0}+1$, where $q=p^r,\ r\geq 0,\ n=p_{a+1},\ p_i=1+q+\cdots+q^{i} $ and $a\geq 2$, $\beta =1-\frac{1}{p_a}(1-\frac{p_a - 1}{q'.p_a})$. It is  the family introduced in \cite{quasi-subfield}.
\item[Type 1bis] $X^{p^{n-1}}+\cdots+X^{p^{2}}+X^{p}+X$, $n'=n-1$, $\beta=1-\frac{1}{(n-1)^2}$.
\item[Type 2] $L_{f_a}$ with $f_a= \left\{\begin{array}{ll}X^{q^d-1}+\cdots+X^{q-1}+
1 & \text{ if } a=0 \\
X^{q^d}+\cdots+X^q+X+a  & \text{otherwise } \\
\end{array}\right.$,  $n = q^{d+1}-1$, $q=p^r$, $r\geq 1$,  $a\in \mathbb{F}_q$, $\beta =  1 - \frac{q^{d-1}}{(1+q+\cdots+q^{d-1})^2} $
\item[Type 3] Inverses of Type 1 and inverses of Type 2.
\end{description}
\end{prop}

\begin{proof}{
Type 1 is proven in \revision{\cite[Lemma~4.3]{quasi-subfield}}.
Moreover it is obvious that $L_{X-1}$ is a quasi-subfield polynomial over $\mathbb{F}_{p^n}$ for any $p$ prime and $n$ (tolerating here $\ell=0$). By Proposition~\ref{prop:inversion}, its inverse  $L_{(X^n-1)/(X-1)} =L_{X^{n-1}+\cdots+X^2+X+1}$ is a quasi-subfield polynomial in $\mathbb{F}_{p^n}$. This proves the Type 1bis. One can notice that it is in fact a particular case of Type 1 (when $r=0$).

For Type 2 polynomials, we need to show that $f_a$ divides $X^n-1$ and thus we look at the factorisation of $X^{q^{d+1}-1}-1$. It appears easier to compute this by looking at $X(X^{q^{d+1}-1}-1)=X^{q{(d+1)}}-X$ since the Frobenius is easy to compute in $\mathbb{F}_{p^n}$. One can observe with $g=X^{q^d}+\cdots+X^q+X$, we have that $X^{q^{d+1}}-X=g^q-g$, Thus, as in Berlekamp's polynomial trace factorization algorithm, we get that $X^{q^{d+1}}-X=\prod_{a\in\mathbb{F}_q} \gcd(X^{q^{d+1}}-X,g+a)$. But we also have $X^{q^{d+1}}-X=(g+a)^q-(g+a)$ so $g_a|X^{q^{d+1}}-X$ and $X^{q^{d+1}}-X=\prod_{a\in\mathbb{F}_q} (g+a)=Xf_0\prod_{a\in\mathbb{F}_q*} f_a$

That is why, $\prod_{a\in\mathbb{F}_q}f_a=X^n-1$.
This gives that Type 2 polynomials split completely  over $\mathbb{F}_{p^n}$. Thus it only remains to verify that $\beta\leq 1 $.

If $a=0$, then $n'=q^{d}-1$ and $\ell=q^{d-1}-1$ hence
\begin{equation*}
\begin{split}
\beta & = \frac{(q^{d-1}-1)(q^{d+1}-1)}{(q^{d}-1)^2} =1-\frac{q^{d+1}+q^{d-1}-2q^{d}}{(q^{d}-1)^2} = 1 - \frac{q^{d-1}(q-1)^2}{(q^d-1)^2} \\
 & = 1 - \frac{q^{d-1}}{(1+q+\cdots+q^{d-1})^2} <1, \\
\end{split}
\end{equation*}
while if $a\neq 0$, then $n=q^{d+1}-1$, $n'=q^{d}$, $\ell=q^{d-1}$ hence $$\beta=\frac{q^{d-1}(q^{d+1}-1)}{q^{2d}}=1-\frac{1}{q^{d+1}}<1.$$

Proposition~\ref{prop:inversion} addresses Type 3 polynomials. We know that for Type 1 we have $Xh^q=h+X^n-1$ thus $X^n-1=h(Xh^{q-1}-1)$ and the inverse of $L_h$ is $L_{Xh^{q-1}-1}$. For Type 2, we have $\prod_{a\in \mathbb{F}_p}f_a=X^n-1$, thus the inverse of $L_{f_a}$ is $L_{\prod_{b\neq a}f_b}$.
}\end{proof}

\revision{Recall that this list does not cover all the equivalence classes. It was only conjectured from what was found with small $n$ and small $p$ and coefficients values in $\left\{0,1,-1\right\}$.} For example, when we launch the algorithm for very small $n$ with coefficients allowed to be anything in $\mathbb{F}_p$, we get for $p=5$ and $n=4$, that $L_{(X^2+X+3)}$ is a linearized QSP. Indeed $(X^2+X+3) (X^2-X+3) =(X^2+3)^2-X^2=X^4+X^2-1-X^2=X^4-1$, so $X^2+X+3$ splits in $\mathbb{F}_{5^4}$ and $\beta = 4 \times 1/4=1$. Moreover, computing its  equivalence class using Proposition~\ref{prop:k_property}, we observe that no element of its equivalence class has all its coefficients in $\left\{0,1,-1\right\}$.

\subsection{Examples of multiplicative quasi-subfield polynomials}

We now study another family of quasi-subfield polynomials considered in~\cite{quasi-subfield}, namely polynomials  whose roots form a multiplicative group of $\mathbb{F}_{p^n}$.

More precisely, we consider quasi-subfield polynomials of the type 
$$L=X^{p^{n'}}-X^{a}$$ 
together with an integer $r$ such that
$a=p^{n'} \mod r$, $r| p^n-1$ and $n'>\log_p(r)$. 
Indeed, $L$ factors as $X^a(X^{p^{n'}-a}-1)$, so the number of roots of $L$ in $\mathbb{F}_{p^n}$ is at most $1+p^{n'}-a$.
  Moreover, there are $\gcd(k,p^n-1)$ roots of $X^k-1$ in $\mathbb{F}_{p^n}$. In order to have the maximal number of distinct roots, we must choose tuples $(p,n,n',r)$ such that for $a=p^{n'}\bmod r$, and $\gcd(p^{n'}-a,p^n-1)=p^{n'}-a$, \revision{\emph{i.e.}} $p^{n'}-a|p^n-1$.

\begin{prop}[Multiplicative quasi-subfield polynomials]
\label{prop:mult_families}
Let $p,n,n',r$ be defined in any of the following three ways:
\begin{enumerate}
\item Let $p$ prime and $k\geq2$ and $i\geq1$ integers. Let $n=2ik$, $n'=i(2k-1)=n-i$ and $r=\frac{p^{n}-1}{p^{2i}-1}$;

\item Let $p=k^n+k-1$ prime and $k\geq 2 $ an integer. Let $n'=1$ and $r=(p-k)/(k-1)$;

\item Let $p=k^n-k-(-1)^n$ be prime, $n>2$ and $k> 1$ integers such that $k^n \gg1$.
Let $n'=n-1$ and let $r=\frac{(p^n-1)(k-(-1)^n)}{(k^{n}-k)(k^n-(-1)^n)}$.
\end{enumerate}
Let $a=p^{n'} \bmod r $ and let $L=X^{p^{n'}}-X^a\in\mathbb{F}_{p^n}[X]$. Then $L$ is a quasi-subfield polynomial.
\end{prop}

We make a few observations before proving this proposition.
For the first family when $p=2$, $i=1$ and $k=2$, we get $r=(2^4-1)/(2^2-1)=5$ and $a=3$, and thus $L=X^8-X^3$ is a quasi-subfield polynomial. On the other hand, $\beta(L) = \log_2(3).4/3^2\simeq 0.70 <0.75 $. This shows that Theorem~\ref{thm:low_bound} is not valid for multiplicative quasi-subfield polynomials.

In the second and last families, we can choose $n$ prime as is the case for the subfield curves recommended by NIST. \revision{However, there are values of $n$ which may not lead to any suitable parameter set for the second and third types. For example, with $n=5$ and $k>1$, the integer $k^5+k-1=(k^3 + k^2 - 1)(k^2 - k + 1)$ so $k^5+k-1$ is not prime.  More generally, for all $n\equiv 5 \mod 6$, $(k^2 -k +1)|(k^n+k-1)$ and thus any integer of the shape $k^{5+6i}+k-1$ with $k>1$ is composite. Similarly, for all $n\equiv 2 \mod 6$, $(k^2-k+1)|(k^n-k+1)$ and thus there is no prime of the shape $k^{2+6i}-k+1$ with $k>1$ and $i>0$.}

The last two families overlap when $n=2$ as $(k-1)^2+(k-1)-1=k^2-k-1$. We excluded the case $n=2$ in the last family, because such a choice of $n'$ and $r$ would lead to $\beta=0$, which is not allowed in our definition of quasi-subfield polynomials. Yet, thanks to the  last two families, we have a multiplicative quasi-subfield polynomial for any $n$ and $p=k^n-k-(-1)^n$ prime.

Finally, it is worth noticing that the case $p=k^n-k-(-1)^n$ is the most promising one among the families introduced. Indeed, primes of the form $f(2^{m})$, where $f(x)$ is a low-degree polynomial with small integer coefficients, are often used in cryptography since they were introduced in \cite{solinas1999generalized}. Indeed as well as for Mersenne primes, they allow fast modular reduction. They are called Solinas primes\index{Solinas primes}, or generalized Mersenne primes. Coming back to our exemple, $f(x)=x^n-x-(-1)^n$ verifies the constraint required about the weights of the coefficients, so the last family when applied with $k$ a power of $2$ corresponds to Solinas primes.  It is then important to notice that Curve448, which is part of the approved elliptic curves for use by the US Federal Government, uses a prime exactly of this shape: $p=2^{448}-2^{224}-1$.\cite{hamburg_ed448-goldilocks_2015}\cite{computer_security_division_transition_2017}. Moreover, four others curves that were recommended by NIST in 1999 \cite{nist1999recommended} also uses Solinas primes:  p-192 ($p=2^{192}-2^64-1$), p-224 ($p=2^{224}-2^{96}+1$) and p-256 ($p=2^{256}-p^{224}+2^{192}+2^{96}-1$) and p-384 ($p=2^{384}-2^{128}-2^{96}+2^{32}-1$). Therefore, it may seem interesting to study more deeply multiplicative quasi-subfield polynomials when $p$ is a Solinas prime. For a list of Solinas primes of the shape $2^n-2^m \pm 1$, one can consult \cite{solinas1999generalized}.  Of course, this approach is still far from threatening the security of these curves: they are defined on a prime field $\mathbb{F}_p$ while we are considering an extension field $\mathbb{F}_{p^n}$ with $n\geq 2$ and have $\beta\simeq 1$ so we obtain a complexity of $O(p^{0.95n})$. (See Remark~\ref{rem:1} for the detail)

\begin{proof}

In each case, we show that for $a:=p^{n'} \bmod r$, we have $p^{n'}-a|p^n-1$ and  $\beta:=\frac{n \log_p{a}}{n'^2}\leq 1$.

\begin{enumerate}
    \item 
 We first note that $r=\frac{p^{n}-1}{p^{2i}-1}$ is an integer since $2i|n$.

We now show that $a= p^{n'} \bmod r=\frac{p^{i(2k-1)}+1}{p^{i}+1}$.
Indeed, we have
$$p^{n'} - \frac{p^{i(2k-1)}+1}{p^{i}+1} 
%=p^{2ik-i} - \frac{p^{2ik-i}+1}{p^{i}+1}
%=\frac{p^{2ik}-1}{p^{i}+1}
=\frac{p^{n}-1}{p^{2i}-1}(p^i-1)
=r(p^i-1)$$
and

\begin{equation*}
    \begin{split}
\frac{p^{i(2k-1)}+1}{p^{i}+1} \frac{1}{r}
=\frac{(p^{i(2k-1)}+1)}{(p^{i}+1)}\frac{(p^{2i}-1)}{(p^{2ik}-1)}
&=\revision{(p^{i(2k-1)}+1)\frac{(p^{i}-1)}{(p^{2ik}-1)}}\\
&=\revision{\frac{p^{2ik}-p^{i(2k-1)}+p^i-1}{(p^{2ik}-1)}}\\
&=1 -\frac{p^{i(2k-1)}-p^i}{p^{2ik}-1} <1
    \end{split}
\end{equation*}
since $2k-1>1$.

Therefore, we have $p^{n'}-a=r(p^i-1) = \frac{p^{2ik}-1}{p^{i}+1} $ and thus $p^{n'}-a|p^n-1$. 
This implies that $L$ splits completely over $\mathbb{F}_{p^n}$ and it has  $p^{n'}-a+1= \frac{p^{2ik}-1}{p^i+1}+1 \approx p^{i(2k-1)} =p^{n'}$ roots.

Moreover, using $a=\frac{p^{i(2k-1)}+1}{p^{i}+1}=\sum_{j=0}^{2k-2}(-p^i)^{j}\leq p^{i(2k-2)}$, we get 
$$\beta= \frac{\log_{p}(a)\cdot n}{n'^2}\leq\frac{i(2k-2)\cdot 2ik}{(i(2k-1))^2}=1-\frac{1}{(2k-1)^2}\leq 1.$$

\item We first note that $r=\frac{p-k}{k-1}=\frac{k^n-1}{k-1}$ is integer since $k-1|k^n-1$. 
Moreover, $p=\frac{p-k}{k-1}(k-1)+k=r(k-1)+k$ with $k<1+k+\cdots+k^{n-1}=r$ so $a=(p \mod r)= k$ and $p-a=k^n-1$.
Therefore,
\begin{equation*}
    \begin{split}
       p^n-1  &=(k^n+k-1)^n-1=\sum_{i=1}^n\binom{n}{i}(k^n-1)^ik^{n-i}  \\
& =(k^n-1)\sum_{i=1}^n\binom{n}{i}(k^n-1)^{i-1}k^{n-i}\\
& =(p-a)\sum_{i=1}^n\binom{n}{i}(k^n-1)^{i-1}k^{n-i}.
    \end{split}
\end{equation*}
Consequently, we have $p-a|p^n-1$, so $L=X^p-X^a$ splits in $\mathbb{F}_{p^n}$ and it has $p^{n'}-a+1=p-a+1=(k^n+k-1)-k+1=k^n$ roots. This is very close to $p=p^{n'}$ if $k^n\gg1$. 
Furthermore, since $k^n\leq k^n+k-1=p$, we have 
$$\beta= \log_p(a)\cdot n/1=\log_p(a^n)=\log_p(k^n)\leq 1.$$

\item  The third proof is very similar to the first two proofs, and presented in \ref{app:appendix_mult_qsp}.
\end{enumerate} 

\end{proof}

\subsection{Quasi-subfield polynomials with \texorpdfstring{$n$}{n} Mersenne}

When $n=2^k-1$ is a Mersenne prime, $(X^n-1)/(X-1)$ has $(n -1)/k$ irreducible factors of degree $k$ over $\mathbb{F}_2$, which gives a large number of potential candidates for linearized quasi-subfield polynomials in $\mathbb{F}_{2^n}$.
We note that for $p=2$, $p^{d+1}-1$ is a Mersenne number\index{Mersenne number}, hence Type 2 of Proposition~\ref{prop:add_families} gives examples of such polynomials.

The case of linearized quasi-subfield polynomials with $n$ a Mersenne prime number is also treated in the appendix  of~\cite{quasi-subfield}. Interestingly, \cite{quasi-subfield} argued that such parameters were unlikely to exist.
We now recall (and slightly extend) their heuristic argument, and we show that Type 2 polynomials from Proposition~\ref{prop:add_families} give a counter-example to it.

\paragraph{Reasoning from~\cite{quasi-subfield}}

Let us consider $k$ such that $n=2^k-1$ is prime, and  denote by $N(k,n')$ the number of distinct polynomials of degree $n'$ that divide $X^n-1$. Then \cite{quasi-subfield} gives the following lemma:
\begin{lem}
We have $N(k,n')=\binom{\lfloor n/k \rfloor}{\lfloor n'/k \rfloor}$ if $ n' \bmod k \in\{0,1\}$, and $N(k,n')=0$ otherwise.
Moreover, $\log\left( \binom{\lfloor n/k \rfloor}{\lfloor n'/k \rfloor}\right)\simeq (n'/k) \log(n/n')$ when $n/k \gg 1$ and $n'/k \gg 1$. 
\end{lem}

The argument in~\cite{quasi-subfield} relies on the following heuristic approximation: for $n$ a Mersenne prime, we may assume that the density of ``sparse enough" polynomials (\emph{i.e.} polynomials of the shape $X^{n'}-\lambda(X) $ with $\deg(\lambda)$ small) is identical
for factors of $X^n- 1$ as for random polynomials of the same degree. 

Since in $\mathbb{F}_2[X]$, there are $2^{n'}$ monic polynomials of degree $n'$ and $2^{\ell}$ monic polynomials of degree at most $\ell$, this assumption allows us to approximate  the number
of polynomials of degree $n'$ that divide $X^n - 1$ and are sparse enough by $N(k,n')2^{\ell-n'}$. Accordingly, such polynomials a priori exist if and only if 
$$\ell> n' -(n'/k)\log(n/n').$$

The case considered in the appendix of the article is when the quasi-subfield polynomial approach beats generic algorithms on ECDLP, which as we will prove in Lemma~\ref{prop:best_alpha} requires $\alpha_{\beta}=\frac{1}{2\calgo \beta}\geq1$ for some algorithmic constant $\calgo $.

We recall their argument in this case first, even if Type 2 does not fall in this category since its $\alpha_{\beta}\simeq \frac{1}{2\calgo } $ is not bigger than 1.
To improve on generic algorithms, we want $\alpha_{\beta} = \frac{1}{2\calgo \ell n/n'^2 }\geq 1$ hence  $\ell\leq \frac{n'^2}{2\calgo n}$.
With the previous constraint on $\ell$, we obtain $\frac{n'^2}{2\calgo n}> n' -(n'/k)\log(n/n')$. Thus, since $k\simeq \log(n)$, we get $\frac{n'}{2\calgo n}> 1 -\log(n/n')/\log(n)= \log(n')/\log(n)$. Therefore, $\frac{\log(n)}{2\calgo n}> \frac{\log(n')}{n'}$.
Since $2\calgo \simeq 10$, and $n'<n$, this inequality can never be satisfied (except if $n'=1$) so according to the above heuristic approximation, there should not be any linearized quasi-subfield polynomials with $n$ Mersenne and big $n$ and $n'$ beating the generic algorithms.

\paragraph{The case of Type 2 polynomials}
The same reasoning can be extended to quasi-subfield polynomials that do not verify $\alpha_{\beta}\geq1$.

In this case, we only require  $\beta=\ell \cdot n/n'^2\leq 1$. Therefore, we have $\ell\leq n'^2/n$ (instead of $\ell\leq \frac{n'^2}{2\calgo n}$ before). This constraint added to the same heuristic as before gives: $\frac{n'^2}{n}> n' -(n'/k)\log(n/n')$  which similarly as in the previous paragraph gives $\frac{\log(n)}{n}> \frac{\log(n')}{n'}$. Since the function $\log(x)/x$ is decreasing for $x>e$, we deduce (following the same heuristic reasoning) that quasi-subfield polynomials are unlikely to exist for $n>n'\geq 3.$

This conclusion, however, is contradicted by the existence of Type 2 polynomials from Proposition~\ref{prop:add_families}.

\paragraph{On the heuristic approximations used in~\cite{quasi-subfield}}   The above contradiction shows that the heuristic approximation used in~\cite{quasi-subfield} idoes not hold in general:  when $n$ is a Mersenne prime, there exists an $\ell$ such that the density of ``sparse enough" polynomials (\emph{i.e.} polynomials of the shape $X^{p^{n'}}-\lambda(X) $ with $ \deg(\lambda)\leq \ell$) is bigger for factors of $X^n- 1$ than for random polynomials.

A similar heuristic in~\cite{quasi-subfield} says that there are only rare parameters for which we can have a quasi-subfield multiplicative polynomials. It uses really similar arguments to the ones introduced before for the case where $n$ is a Mersenne prime. Property~\ref{prop:mult_families} shows that this heuristic  about the distribution of completely splitting polynomials also fails.

\section{Application to Cryptography\label{sec:applications}}

While quasi-subfield polynomials are mathematical objects of independent interest, the main motivation for their introduction in~\cite{quasi-subfield} is a cryptographic application.
In this section we first recall the Elliptic Curve Discrete Logarithm Problem (ECDLP) and standard approaches to solve it. We then describe  Huang \emph{et~al.}'s algorithm~\cite{quasi-subfield} using quasi-subfield polynomials and we explain how its complexity crucially depends on the parameter $\beta$ of the polynomial. \revision{Next, we} apply our results to this ECDLP algorithm, and discuss the resulting complexity. \revision{Finally, we introduce some aspects of coding theory where our results on linearized polynomials could be useful}.

\subsection{ECDLP and previous ECDLP algorithms}

Let us consider an ECDLP instance: Let $\mathcal{E}$ be an elliptic curve on $K=\mathbb{F}_{p^n}$, $P$ a point on the curve $\mathcal{E}$ and $Q$ a point in $<P>$, the group generated by $P$. We are looking for $k$ such that  $Q=kP$\index{ECDLP instance}.

Before considering the algorithm using QSPs \cite{quasi-subfield},
we recall two algorithms for solving the ECDLP and their complexity.
\begin{itemize}
\item Exhaustive search (or brute-force algorithms) : it corresponds to the computation all the elements of $<P>$ until finding $Q$. The cost is $O(<P>)=O\left(p^n\right)$ for typical parameters.
\item Generic algorithms such as Baby-Step-Giant-Step or Pollard-Rho~\cite{jofc-1999-14267}: The complexity is $O(\sqrt{|<P>|})\approx O\left(p^{n/2}\right)$. \index{Generic algorithm}
\end{itemize}

These will be used as benchmarks to assess the performance of our algorithm.
For more information about these algorithms and other approaches to solve the ECDLP, the reader can consult \textit{Recent progress on the elliptic curve discrete logarithm problem}~\cite{Galbraith2016} by Galbraith and Gaudry. It is also worth noticing that these two algorithms can solve the discrete logarithm problem in any group and we can hope that the new algorithm, which uses the structure of the group, has a better complexity.

When $n$ is composite, we can write $n=\tilde{n}k$ and consider $q=p^k$ so that $\mathbb{F}_{p^n}=\mathbb{F}_{q^{\tilde{n}}}$. Better algorithms exist in this situation: Gaudry \cite{gaudry_index_2009} succeeded in 2009 to find an algorithm solving the elliptic curve discrete logarithm problem on $\mathbb{F}_{q^{\tilde{n}}}$ in $O(q^{2-2/\tilde{n}})$.  
For $\tilde{n}=2$, it leads to an algorithm with cost $O(p^{(n/2)(2-1)})=O(p^{n/2})$ comparable with generic algorithms. For $\tilde{n}=3$, it leads to an algorithm with cost $O(p^{(n/3)(2-2/3)})=O(p^{4/9n})$, slightly better than generic algorithms.
Diem also proved that there exists a sequence of prime
powers $Q_i = q^{n_i}_i$ with $n_i \simeq \sqrt{log(q_i )}$ such that the ECDLP in $\mathcal{E}(F_{Q_i} )$ can be solved in subexponential time \cite{diem_discrete_2011}. It works
with any elliptic curve over $\mathbb{F}_{Q_i}$ and uses an approach similar to the one introduced below but with subfield polynomials instead of quasi-subfield polynomials.
This was one of the motivation of this new approach.

\subsection{The quasi-subfield approach}
We will now introduce the algorithm of \cite{quasi-subfield}, which uses quasi-subfield polynomials to solve elliptic curve discrete logarithm problems.

Let $\mathcal{E}$ be an elliptic curve on $K=\mathbb{F}_{p^n}$, $P\in \mathcal{E}$ and $Q\in< P >$. The elliptic curve discrete logarithm problem asks for computing $k$ such that  $Q=kP$. 
For simplicity and concreteness, we assume the curve is given in reduced Weierstrass coordinates.

Let $R\in\mathbb{F}_{p^n}[X]$ be a quasi-subfield polynomial. We define $V$ as the set of the roots of $R$ and $\mathcal{F}:=\left\{ (x,y)\in \mathcal{E} |x\in V\right\} $.
The algorithm first computes more than $|V|$ \emph{relations} of the shape: 
$$a_jP+b_jQ =P_1+\cdots+P_m$$ 
with $a_j,b_j$ random and $P_i\in\mathcal{F}$. Linear algebra on the relations then gives the value of $k$ such that $Q=kP$. 

In order to compute these relations, the algorithm uses  Semaev's summation polynomials~\cite{semaev_summation_2004}: \index{Semaev summation polynomials}
for an elliptic curve $\mathcal{E}$ defined over a field $K$,  the $r$th summation polynomial $S_r\in K[X]$ is such that $$S_r(x_1,\cdots,x_r)=0 \Leftrightarrow \exists (x_1,y_1),\cdots (x_r,y_r)\in \mathcal{E},(x_1,y_1)+\cdots+(x_r,y_r)=0$$

For given $a_j,b_j$, we compute $a_jP+b_jQ=(X_j,Y_j)$. 
Then, computing $P_1,\cdots,P_m$ such that 
$$a_jP+b_jQ = P_1+\cdots+P_m=(x_1,y_1)+\cdots+(x_m,y_m)$$
with $x_i\in V$ amounts to finding $x_1,\cdots,x_m \in V$ such that 
$S_{m+1}(X_i,x_1,\cdots,x_m)=0$ and then finding the associated $y_i$.

The polynomial equation $S_{m+1}(X_i,x_1,\cdots,x_m)=0$ is solved as follows.
Let $\mathcal{M}$ be the set of monomials in $K[x_1,\cdots,x_m]$, and let $i$ be a positive integer.
For $f=\sum_{M\in\mathcal{M}}a_MM\in K[x_1,\cdots,x_m] $, let $F^{i}(f)=\sum_{M\in\mathcal{M}}a_M^{p^i}M$.
Let also $\phi: K[x_1,\cdots,x_m] \rightarrow K[x_1,\cdots,x_m]$ defined by
$$f(x_1,\cdots,x_m) \mapsto F^{n'}(f)(\lambda(x_1),\cdots,\lambda(x_m)).
$$
Note that $f^{p^{n'}} \equiv \phi(f) \bmod (x_1^{p^{n'}}- \lambda(x_1),\cdots, x_m^{p^{n'}}- \lambda(x_m))$.

Finally, let $S^{(0)}(x_1,\cdots,x_m)=S_{m+1}(X_i,x_1,\cdots,x_m)$
and for $k\in\left\{1,\cdots,m-1\right\}$, let $S^{(k)}(x_1,\cdots,x_m)=\phi(S^{(k-1)}(x_1,\cdots,x_m))$.
One can then solve the polynomial equation $S_{m+1}(X_i,x_1,\cdots,x_m)=0$ by solving the system $\mathcal{S}=\left\{S^{(k)}=0\right\}_{k=1}^{m-1}$. This is a sparse polynomial system with $m$ equations and $m$ variables.

\subsection{Complexity of the quasi-subfield approach}

We now recall the complexity estimations of this algorithm as given in~\cite{quasi-subfield}.

We know that $\mathcal{S}=\left\{S^{(k)}\right\}_{k=1}^{m-1}$ is a sparse polynomial system of $m$ equations and $m$ variables. Therefore, it can be solved efficiently using Rojas' sparse resultant algorithm \cite{rojas_solving_1999} and a univariate polynomial root finding algorithm such as BTA \cite{berlekamp1968algebraic}. According to~\cite{quasi-subfield} (Lemma 3.1) the cost of this step is $\tilde{O}(m^{5.188}(3p^{\ell})^{\calgo m^2})$. Here we introduce the notation $\calgo $ as the numerical value 4.876 used in~\cite{quasi-subfield} may be suboptimal. 
Moreover, the system has solutions only with probability $\frac{|\mathcal{F}|^m/m!}{p^n}$ since $(X_i,Y_i)$ is a random point on $\mathcal{E}$ with $|\mathcal{E}|\simeq p^n$ and the number of sums of $m$ points in $\mathcal{F}$ is approximately $|\mathcal{F}|^m/m!$. Also, heuristically, half of the
values in $V$ are the x-coordinates of exactly two points on the curve so $|\mathcal{F}|\simeq |V|\simeq p^{n'}$.
As we need  $p^{n'}$ relations of this type, the cost of the relation search phase is $p^{n'}\frac{m!p^n}{p^{n'm}}\tilde{O}(m^{5.188}(3p^{\ell})^{\calgo m^2})$.

Once all the $p^{n'}$ relations are gathered, each of them involves $m$ points. Therefore, the system built from these relations is sparse. Thus, a sparse linear algebra algorithm can be used to finish the computation \cite{wiedemann_solving_1986}, at a cost approximately  $mp^{2n'}$. This gives the complete cost of the algorithm:
$$m!p^{n-n'm+n'}\tilde{O}(m^{5.188}(3p^{\ell})^{\calgo m^2})+mp^{2n'}$$
Rewriting this expression to make $\beta$ appear, we get the following estimation of the complexity:

\begin{prop}[Complexity of Huang \emph{et~al.}'s algorithm]
Let $P=X^{p^{n'}}-\lambda(X)$ be a $\beta$-quasi-subfield polynomial  and let $\ell = \log_p(\deg \lambda)$. If  $|\mathcal{F}|\simeq|\mathcal{V}|\simeq p^{n'}$, the complexity of Huang \emph{et~al.}'s algorithm is  $$\tilde{O}\left(m!p^{n\left(1+\calgo \beta \left(\frac{n'm}{n}\right)^2 -\frac{n'm}{n}\right)+n'}m^{5.188}3^{\calgo m^2}\right)+mp^{2n'}$$ 
where $\calgo $ is a constant involved in the cost of the resolution of the system $\mathcal{S}$ currently majored by 4.876.
\end{prop}

In the following we define $\alpha=n'm/n>0$, and we try to find $\alpha$ which minimises the complexity.
 We will assume that $m$ is fixed. 

\begin{prop}[Best choice of parameters]
\label{prop:best_alpha}
We can observe the following results in order to optimize the complexity: 
\begin{itemize}
\item The minimal complexity is obtained for $\alpha=\alpha_{\beta}$ with $\alpha_{\beta}:=\frac{1}{2\calgo \beta}$. Then, the  complexity becomes   $\tilde{O}\left(p^{\max\left(2\alpha_{\beta}/m,1-\alpha_{\beta}(1/2-1/m)\right)n}\right)$.
\item In order to beat brute force algorithms,  $m>\max(2\alpha_{\beta},2)$ is required. So we have interest not to choose a very small integer for $m$.
\item If $\alpha_{\beta}<2$ and $m\gg1$, then the complexity becomes $\tilde{O}\left(p^{(1-\alpha_{\beta}/2)n}\right)$
\item Therefore, to beat generic algorithms, we need $\alpha_{\beta}>1$
\end{itemize}
\end{prop}

We remark that the condition $\alpha_{\beta}<2$ is not really restrictive. Indeed for all the quasi-subfield polynomials exhibited in this paper, we have $\alpha_{\beta}<\alpha_{0.5}<1$.

\begin{proof}
Let us now prove these four results.
The complexity of the algorithm is bounded by  $\tilde{O}(m!p^{n(1+\calgo \beta (n'm/n)^2 -n'm/n)+n'}m^{5.188}3^{\calgo m^2})+mp^{2n'}$ which with the $\alpha$-notation and the fact that $m$ is considered as a fixed integer, can be rewritten as 
$\tilde{O}(p^{n(\calgo \beta\alpha^2 -\alpha+1)+\alpha n/m}+p^{2\alpha n/m})$.

Since $\calgo \beta\alpha^2 -\alpha+1$ is minimum for $\alpha = \frac{1}{2\calgo \beta}=\alpha_{\beta}$ (we recall that we only consider $\beta>0$) and then has minimal value $\calgo \beta \frac{1}{(\calgo \beta)^2}-\frac{1}{2\calgo \beta}+1=1-\frac{1}{4\calgo \beta }=1-\frac{\alpha_{\beta}}{2}$, we get that the complexity can be rewritten as $$\tilde{O}\left(p^{\max\left(2\alpha_{\beta}/m,1-\alpha_{\beta}(1/2-1/m)\right)n}\right).$$

In order to beat the brute force algorithms (which corresponds to a complexity of $O(p^n)$), what we need is to have on one side  $2\alpha_{\beta}/m<1$ which is true as soon as $m > 2\alpha_{\beta}$, and on the other side, $1-\alpha_{\beta}(1/2-1/m)<1 $, namely $m>2$.	Hence $m>\max(2\alpha_{\beta},2)$.

Moreover, one can notice that $2\alpha_{\beta}/m\leq 1-\alpha_{\beta}(1/2-1/m)$ if only if $\alpha_{\beta}(1/m+1/2)\leq 1 $. Therefore if $m\gg1$ then $\alpha_{\beta}\leq 2$ implies $2\alpha_{\beta}/m \leq 1-\alpha_{\beta}((1/2-1/m)$, so we can rewrite the complexity as $\tilde{O}\left(p^{(1-\alpha_{\beta}/2)n}\right)$.

Generic algorithms have a complexity of $O(p^{n/2})$, therefore we need $\alpha_{\beta}>1$ to run faster than them.
\end{proof}

The following table gives concrete complexity estimates for various values of $\beta$, assuming $\calgo =4.876$.

\begin{figure}[H]
\centering
$$
\begin{array}{|c|c|c|c|c|c|c|c|}
	\hline
	\beta & 1.0 & 0.8 &  0.6 & 0.4 & 0.2 & 0.15  & 0.1 \\ \hline
 1-\alpha_{\beta}/2 &  0.949 & 0.936  & 0.915 & 0.872 & 0.744  & 0.658  & 0.487 \\ \hline
\end{array}
$$
\caption*{Complexity estimates of Huang \emph{et~al}'s algorithm for various values of $\beta$. By Proposition~\ref{prop:best_alpha} the complexity of Huang \emph{et~al.}'s algorithm is $\tilde{O}\left(p^{(1-\alpha_{\beta}/2)n}\right)$. 
}

\end{figure}

\begin{remark}
We observe that for $\beta=1$, we have $1-\alpha_{\beta}/2\approx 0.95$ so we get a complexity slightly better than the one  of brute force algorithms. \label{rem:1} 
We can beat generic algorithms for $\alpha_{\beta}>1$, which for this specific value of $\calgo $ implies $\beta<0.103$.
\end{remark}

\subsection{Impact of our results on ECDLP}

We will now study the consequences of Theorem~\ref{thm:low_bound}.
Let $L$ be a linearized quasi-subfield polynomial. Then by Theorem~\ref{thm:low_bound} we have $\beta(L)\geq3/4$ and with 
$\calgo \simeq 4.876$ we get
$\alpha_{\beta}=\dfrac{1}{2\cdot \calgo \beta(L)}\leq \dfrac{2}{3\cdot \calgo }< 1/7$. 
This shows that $a_{\beta}< 1$ and it is not possible to beat generic algorithms with $L$. The best complexity we can hope is indeed $\tilde{O}(p^{(1-\alpha_{\beta}(1/2-1/m))n})$, which is bigger that $\tilde{O}(p^{(1-1/14)n})$.

The previous estimation uses  the approximation  $\calgo \simeq 4.876$. If we succeeded to have $\calgo <1.5$ then, when $\beta(L)= 3/4$, we would have $\alpha_{\beta}>1$, so such a polynomial $L$ could allow us  to have an algorithm running faster than generic algorithms.

All the quasi-subfield polynomials exhibited in this article have $\beta >0.7$ and thus $\alpha_{\beta}<1$. In particular, none of them currently leads to an algorithm running faster than generic algorithms. 

\revision{
\subsection{Links with coding theory}

Linearized polynomials have attracted considerable interest, and  our results can therefore be used in other contexts as well.

For example, linearized polynomials occur in rank-metric codes. The characterisation of completely splitting linearized trinomials given in  \cite{santonastaso2020linearized}  is used  in the same article to study codes of the shape $ C_{3,n,\sigma} = \langle x, x^{\sigma}, x^{\sigma^3} \rangle_{\mathbb{F}_{q^n}}$, with both their result on existence and non existence of such trinomials being used.
In \cite{csajbok2020mrd}, codes of the shape 
$$\mathcal{C}_T:= \left\{a_0 X^{q^{t_0}} + a_1 X^{q^{t_1}} + \dots a_{k-1}X^{q^{t_{k-1}}}, ~a_0, a_1, \dots ,a_{k-1} \in \mathbb{F}_{q^n}\right\}$$
for sets $T = \left\{t_0 <t_1 < \dots < t_{k-1}\right\}\subset\left\{0,\dots, n-1\right\}$ are studied. Maybe, Theorem~\ref{thm:low_bound}, which gives wider results than \cite{santonastaso2020linearized} on completely splitting linearized polynomials could help to study such codes.

This notion also appears in cyclic subspace codes. For instance in \cite{santonastaso2020linearized}, families of cyclic subspace codes are exhibited via linearized polynomials. Interestingly, the paper uses a parameter called \emph{gap} which characterizes the sparsity of a polynomial and implies bounds on the minimal distance of an associated code.
 The gap is defined as $n'-\ell$ for a linearized polynomial $P=X^{p^{n'}}-\lambda(X)\in\mathbb{F}_{p^n}[X]$ with $\lambda$ of degree $p^{\ell}$.  It is therefore similar to our parameter $\beta:=\dfrac{\ell\cdot n}{n'^2}$.
}

\section{Conclusion \label{sec:conclusion}}

We studied the existence of quasi-subfield polynomials (QSP) introduced by Huang \emph{et~al.}~\cite{quasi-subfield}. 
We proved a new lower bound on the $\beta$ parameter of \emph{linearized} QSP, and we introduced several new QSP families. \revision{We leave as an open problem the classification of all the QSP.
}

The main motivation underlying~\cite{quasi-subfield} is a new algorithm to solve the elliptic curve discrete logarithm problem, with a complexity depending on the $\beta$ parameter of the QSP used. We showed that this algorithm is currently outperformed by other algorithms even with our new QSP families. Moreover, our new bound suggests that Huang \emph{et~al.}'s algorithm will remain worse if only linearized QSPs are used.

\revision{\section*{Acknowledgements}
We would like to thank the reviewers for their helpful comments, especially for pointing out the link between this work and coding theory.}

\bibliographystyle{elsarticle_num}
\bibliography{main.bib}

\begin{thebibliography}{10}
\expandafter\ifx\csname url\endcsname\relax
  \def\url#1{\texttt{#1}}\fi
\expandafter\ifx\csname urlprefix\endcsname\relax\def\urlprefix{URL }\fi
\expandafter\ifx\csname href\endcsname\relax
  \def\href#1#2{#2} \def\path#1{#1}\fi

\bibitem{quasi-subfield}
M.-D. Huang, M.~Kosters, C.~Petit, S.~L. Yeo, Y.~Yun, Quasi-subfield
  polynomials and the elliptic curve discrete logarithm problem, Journal of
  Mathematical Cryptology 14~(1) (2020) 25--38.

\bibitem{mcguire2020some}
G.~McGuire, D.~Mueller, Some results on linearized trinomials that split
  completely, Finite Fields and their Applications (2020) 149.

\bibitem{mcguire2019characterization}
G.~McGuire, J.~Sheekey, A characterization of the number of roots of linearized
  and projective polynomials in the field of coefficients, Finite Fields and
  Their Applications 57 (2019) 68--91.

\bibitem{csajbok2019characterization}
B.~Csajb{\'o}k, G.~Marino, O.~Polverino, F.~Zullo, A characterization of
  linearized polynomials with maximum kernel, Finite Fields and Their
  Applications 56 (2019) 109--130.

\bibitem{berlekamp1968algebraic}
E.~Berlekamp, Algebraic coding theory, World Scientific, 1968.

\bibitem{chen_combinatorial_1996}
W.~Y.~C. Chen, J.~D. Louck,
  \href{http://www.sciencedirect.com/science/article/pii/0024379595901639}{The
  combinatorial power of the companion matrix}, Linear Algebra and its
  Applications 232 (1996) 261--278.
\newblock \href {http://dx.doi.org/10.1016/0024-3795(95)90163-9}
  {\path{doi:10.1016/0024-3795(95)90163-9}}.
\newline\urlprefix\url{http://www.sciencedirect.com/science/article/pii/0024379595901639}

\bibitem{santonastaso2020linearized}
P.~Santonastaso, F.~Zullo, Linearized trinomials with maximum kernel, arXiv
  preprint arXiv:2012.14861.

\bibitem{solinas1999generalized}
J.~A. Solinas, et~al., Generalized mersenne numbers, Citeseer, 1999.

\bibitem{hamburg_ed448-goldilocks_2015}
M.~Hamburg, \href{https://eprint.iacr.org/2015/625}{Ed448-{Goldilocks}, a new
  elliptic curve}, Tech. Rep. 625 (2015).
\newline\urlprefix\url{https://eprint.iacr.org/2015/625}

\bibitem{computer_security_division_transition_2017}
I.~T.~L. Computer Security~Division,
  \href{https://csrc.nist.gov/News/2017/Transition-Plans-for-Key-Establishment-Schemes}{Transition
  {Plans} for {Key} {Establishment} {Schemes} {\textbar} {CSRC}} (Oct. 2017).
\newline\urlprefix\url{https://csrc.nist.gov/News/2017/Transition-Plans-for-Key-Establishment-Schemes}

\bibitem{nist1999recommended}
NIST, Recommended elliptic curves for federal government use (1999).

\bibitem{jofc-1999-14267}
P.~C. van Oorschot, M.~J. Wiener, Parallel collision search with cryptanalytic
  applications, J. Cryptology 12 (1999) 1--28.
\newblock \href {http://dx.doi.org/10.1007/PL00003816}
  {\path{doi:10.1007/PL00003816}}.

\bibitem{Galbraith2016}
S.~D. Galbraith, P.~Gaudry,
  \href{http://link.springer.com/10.1007/s10623-015-0146-7}{Recent progress on
  the elliptic curve discrete logarithm problem}, Designs, Codes and
  Cryptography 78~(1) (2016) 51--72.
\newblock \href {http://dx.doi.org/10.1007/s10623-015-0146-7}
  {\path{doi:10.1007/s10623-015-0146-7}}.
\newline\urlprefix\url{http://link.springer.com/10.1007/s10623-015-0146-7}

\bibitem{gaudry_index_2009}
P.~Gaudry,
  \href{https://linkinghub.elsevier.com/retrieve/pii/S074771710800182X}{Index
  calculus for abelian varieties of small dimension and the elliptic curve
  discrete logarithm problem}, Journal of Symbolic Computation 44~(12) (2009)
  1690--1702.
\newblock \href {http://dx.doi.org/10.1016/j.jsc.2008.08.005}
  {\path{doi:10.1016/j.jsc.2008.08.005}}.
\newline\urlprefix\url{https://linkinghub.elsevier.com/retrieve/pii/S074771710800182X}

\bibitem{diem_discrete_2011}
C.~Diem,
  \href{https://www.cambridge.org/core/product/identifier/S0010437X10005075/type/journal_article}{On
  the discrete logarithm problem in elliptic curves}, Compositio Mathematica
  147~(1) (2011) 75--104.
\newblock \href {http://dx.doi.org/10.1112/S0010437X10005075}
  {\path{doi:10.1112/S0010437X10005075}}.
\newline\urlprefix\url{https://www.cambridge.org/core/product/identifier/S0010437X10005075/type/journal_article}

\bibitem{semaev_summation_2004}
I.~Semaev, \href{http://eprint.iacr.org/2004/031}{Summation polynomials and the
  discrete logarithm problem on elliptic curves}, Tech. Rep. 031 (2004).
\newline\urlprefix\url{http://eprint.iacr.org/2004/031}

\bibitem{rojas_solving_1999}
J.~M. Rojas,
  \href{http://www.sciencedirect.com/science/article/pii/S0747717198902711}{Solving
  {Degenerate} {Sparse} {Polynomial} {Systems} {Faster}}, Journal of Symbolic
  Computation 28~(1) (1999) 155--186.
\newblock \href {http://dx.doi.org/10.1006/jsco.1998.0271}
  {\path{doi:10.1006/jsco.1998.0271}}.
\newline\urlprefix\url{http://www.sciencedirect.com/science/article/pii/S0747717198902711}

\bibitem{wiedemann_solving_1986}
D.~Wiedemann, Solving sparse linear equations over finite fields, IEEE
  Transactions on Information Theory 32~(1) (1986) 54--62.
\newblock \href {http://dx.doi.org/10.1109/TIT.1986.1057137}
  {\path{doi:10.1109/TIT.1986.1057137}}.

\bibitem{csajbok2020mrd}
B.~Csajbok, G.~Marino, O.~Polverino, Y.~Zhou, Mrd codes with maximum
  idealizers, Discrete Mathematics 343~(9) (2020) 111985.

\end{thebibliography}

\appendix

\section{Adaptation of the results from \cite{chen_combinatorial_1996}}
\label{app:powers_of_companion}

\revision{In \cite{chen_combinatorial_1996}, Chen and Louck give a formula for computing the powers of the following companion matrices : 
$$C(u_1,\cdots,u_m):=\begin{bmatrix}
u_1 & u_2 &\cdots & \cdots& u_m \\
1 & 0 &\cdots & \cdots & 0 \\
0 &1 &\cdots & \cdots  & 0 \\
\vdots & \vdots & \ddots  & &\vdots \\
0 & 0 &\cdots &   1 &0 \\
\end{bmatrix}$$
Comparing it with our definition of companion matrices, $$D(a_1,a_2,...,a_m)=
\begin{bmatrix}
0 & 0 &\cdots & \cdots& a_1 \\
1 & 0 &\cdots & \cdots & a_2 \\
0 &1 &\cdots & \cdots  & \vdots \\
\vdots & \vdots & \ddots  & \vdots \\
0 & 0 &\cdots &   1 & a_m\\
\end{bmatrix},$$ we notice that $ D(a_1,a_2,...,a_m)$ is the antitranspose of $C(a_m,\cdots, a_2,a_1)$. 
Formally, we have : $$D(a_1,a_2,...,a_m) = P\cdot C(a_m,\cdots, a_2,a_1)^{T}\cdot P \text{ with }  P=\begin{bmatrix} 0 & \cdots & 0 & 1 \\
                   \vdots & \cdots  & 1 &0 \\
                   \vdots && &  \vdots\\
                   1 & \cdots &  & 0 \end{bmatrix}.$$ 

Since $P^2=Id$, we get that for all $n\geq 0$, $D(a_1,a_2,...,a_m)^n$ is the antitranspose of $C(a_m,\cdots, a_2,a_1)^n$.
Chen and Louck give the following formula for the coefficient $(i,j)$ of $C(u_1,\cdots,u_m)^n$:
$$c_{i,j}^{(n)} = \sum_{k_1,...,k_m} \dfrac{k_j+k_{j+1}+\cdots+k_m}{k_1+\cdots +k_m} \binom{k_1,k_2,\cdots,k_m}{k_1+\cdots+k_m} u_1^{k_1}\cdots u_m^{k_m}$$ where the summation is over non-negative integers satisfying $\sum \ind k_{\ind} = n-i+j$. Moreover, when the previous sum is not defined, \emph{i.e.} when $n=i-j$, $c_{i,j}^{(n)}=1$.

Since applying the antitranspose boils down to swapping the coefficients $(i,j)$ and $(m+1-j,m+1-i)$\footnote{We number the rows, as well as the columns from 1 to $m$.}, we get the expression of the coefficient $(i,j)$ of $D(a_1,a_2,...,a_m)^n$:

\begin{align*}
 d_{i,j}^{(n)} &= c_{m+1-j,m+1-i}^{(n)} \\
 &= \sum_{k_1,...,k_m} \dfrac{k_{m-i+1}+k_{m-i+2}+\cdots+k_m}{k_1+\cdots +k_m} \binom{k_1,k_2,\cdots,k_m}{k_1+\cdots+k_m} a_m^{k_1}\cdots a_1^{k_m}  
\end{align*}
where the summation is over non-negative integers satisfying $$\sum \ind k_{\ind} = n-(m-j+1)+m-i+1=n-i+j$$

Moreover, when the previous sum is not defined (\emph{i.e.} when $n=j-i$) then $d_{i,j}^{(n)}=1$.

In the proof of Lemma~\ref{lem:non_null_coeff}, we consider  $M:=D(\mysquare,\dots,\mysquare,\mypow,0,\dots,0)$ a matrix of dimension $n'$, which leads to 
$M^{n}_{i,j}=1$ if $n=i-j$,
and 
$$M^{n}_{i,j}=\sum\limits_{\substack{\textbf{k}=(k_1,\cdots,k_{n'})\\ \sum\limits_{\ind=1}^{n'}\ind k_{\ind}=n -i+j}}w_{\textbf{k}}\cdot
0^{k_1+\cdots+k_{n'-\ell-1}} \cdot (\mypow)^{k_{n'-\ell}}\cdot \mysquare^{k_{n'-\ell+1}+\cdots+k_{n'}} $$
otherwise, where $\textbf{k}=(k_{\ind})_{1\leq \ind\leq n'}$ are non-negative integers and $$w_{\textbf{k}}=\dfrac{k_{n'-i+1}+\cdots+k_{n'}}{k_{1}+\cdots+k_{n'}}\binom{k_{1}+\cdots+k_{n'}}{k_{1},\cdots,k_{n'}}.$$ }

\section{Some linearized QSP}

\revision{In this section we provide the list of linearized QSP found through the systematic search described in Section~\ref{sec:systematic_search}.
Recall that this search only covers representatives of equivalence classes for $p\in\{2,3,5,7\}$, $n'\leq16$, and coefficients values restricted to $\left\{0,1,-1\right\}$.

For the sake of readability, we list polynomials $f$ instead of their corresponding quasi-subfield polynomials $L_f$. We provide the values of $n$ and $p$ such that $L_f$ is in $\mathcal{Q}^{L}_{p,n}$ and indicate the associated value $\beta$. We mark by a checkmark in the table when the linearized polynomial belongs to the category (as defined in Proposition~\ref{prop:add_families}), except for the last category where we give the value of the inverse.}

	\begin{landscape}
		\begin{longtable}{|C{8cm}|C{0.5cm}|C{0.5cm}|C{1cm}|C{0.4cm}|C{0.4cm}|C{2.3cm}|}
			\hline
			$f$ &  $n$ & $\beta$ & $p$ & T1 & T2  & T3 \\
			\hline
			\endfirsthead
			\hline
			$f$ &  $n$ & $\beta$ & $p$ & T1 & T2  & T3 \\
			\hline
			\endhead
			\hline 
			\endfoot
			\endlastfoot
			$X^{2}+X+1$&3&0.75&2&$\checkmark$&$\checkmark$& \\
			$X^{2}+X+1$&3&0.75&3,5,7&$\checkmark$& & \\
			$X^{3}+X+1$&7&0.78&2&$\checkmark$&$\checkmark$& \\
			$X^{3}+X+1$&8&0.8&3& &$\checkmark$& \\
			$X^{3}+X^{2}+X+1$&4&0.8&2,3,5,7&$\checkmark$& & \\
			$X^{4}+X+1$&15&0.9&2& &$\checkmark$& \\
			$X^{4}+X+1$&13&0.8&3&$\checkmark$& & \\
			$X^{4}+X^{2}+X+1$&7&0.8&2&$\checkmark$&$\checkmark$&$X^3 + X + 1$\\
			$X^{4}+X^{3}+X^{2}+X+1$&5&0.9&2,3,5,7&$\checkmark$& & \\
			$X^{5}+X+1$&21&0.8&2&$\checkmark$& & \\
			$X^{5}+X+1$&24&0.9&5& &$\checkmark$& \\
			$X^{5}+X^{4}+X^{3}+X^{2}+X+1$&6&0.9&2,3,5,7&$\checkmark$& & \\
			$X^{5}-X^{3}-X^{2}+X-1$&8&0.9&3& & &$X^3 + X + 1$\\
			$X^{6}+X+1$&31&0.8&5&$\checkmark$& & \\
			$X^{6}+X^{5}+\cdots+X^{2}+X+1$&7&0.9&2,3,5,7&$\checkmark$& & \\
			$X^{7}+X+1$&48&0.9&7& &$\checkmark$& \\
			$X^{7}+X^{3}+X+1$&15&0.9&2&$\checkmark$&$\checkmark$& \\
			$X^{7}+X^{6}+\cdots+X^{2}+X+1$&8&0.9&2,3,5,7&$\checkmark$& & \\
			$X^{8}+X+1$&63&0.9&2& &$\checkmark$& \\
			$X^{8}+X+1$&57&0.8&7&$\checkmark$& & \\
			$X^{8}+X^{4}+X^{2}+X+1$&15&0.9&2&$\checkmark$&$\checkmark$&$X^7 + X^3 + X + 1$\\
			$X^{8}+\cdots+X^{2}+X+1$&9&0.9&2,3,5,7&$\checkmark$& & \\
			$X^{9}+X+1$&73&0.9&2&$\checkmark$& & \\
			$X^{9}+X+1$&80&0.9&3& &$\checkmark$& \\
			$X^{9}+X^{3}+X+1$&26&0.9&3& &$\checkmark$& \\
			$X^{9}-X^{6}-X^{5}+X^{3}-X^{2}+X-1$&13&0.9&3& & &$X^4 + X + 1$\\
			$X^{9}+\cdots+X^{2}+X+1$&10&0.9&2,3,5,7&$\checkmark$& & \\
			$X^{10}+X+1$&91&0.9&3&$\checkmark$& & \\
			$X^{10}+\cdots+X^{2}+X+1$&11&0.9&2,3,5,7&$\checkmark$& & \\
			$X^{11}+X^{8}+X^{7}+X^{5}+X^{3}+X^{2}+X+1$&15&0.9&2&$\checkmark$& &$X^4 + X + 1$\\
			$X^{11}+\cdots+X^{2}+X+1$&12&0.9&2,3,5,7&$\checkmark$& & \\
			$X^{12}+\cdots+X^{2}+X+1$&13&0.9&2,3,5,7&$\checkmark$& & \\
			$X^{13}+X^{4}+X+1$&40&0.9&3&$\checkmark$& & \\
			$X^{13}+\cdots+X^{2}+X+1$&14&0.9&2,3,5,7&$\checkmark$& & \\
			$X^{14}+\cdots+X^{2}+X+1$&15&0.9&2,3,5,7&$\checkmark$& & \\
			$X^{15}+X^{7}+X^{3}+X+1$&31&0.9&2&$\checkmark$&$\checkmark$& \\
			$X^{15}+X^{14}+\cdots+X^{2}+X+1$&16&0.9&2,3,5,7&$\checkmark$& & \\
			$X^{16}+X+1$&255&0.9&2& &$\checkmark$& \\
			$X^{16}+X^{4}+X+1$&63&0.9&2& &$\checkmark$& \\
			$X^{16}+X^{8}+X^{4}+X^{2}+X+1$&31&0.9&2&$\checkmark$&$\checkmark$&$X^{15} + X^7 + X^3 + X + 1$\\
			$X^{16}+X^{12}+X^{11}+X^{8}+X^{6}+X^{4}+X^{3}+X^{2}+X+1$&21&0.9&2&$\checkmark$& &$X^5 + X + 1$\\
			$X^{16}+\cdots+X^{2}+X+1$&17&0.9&2,3,5,7&$\checkmark$& & \\
			& \vdots &  & &  && \\
			\hline
			\caption{Classification of the outputs of the algorithm \label{tab:linear_qsp}
			}
		\end{longtable}
	\end{landscape}

\section{Proof of the third family of multiplicative QSPs}
\label{app:appendix_mult_qsp}

We will now demonstrate that the third family of Proposition \ref{prop:mult_families} is a family of multiplicative QSPs. We recall that it is defined as $X^{p^{n'}}-X^a \in \mathbb{F}_{p^n}$ with 
\begin{itemize}
    \item $p=k^n-k-(-1)^n$ prime, $n>2$ and $k> 1$ integers such that $k^n \gg1$.
   \item  $n'=n-1$ \item $r=\frac{(p^n-1)(k-(-1)^n)}{(k^{n}-k)(k^n-(-1)^n)}$ 
   \item $a:= p^{n'} \mod r$
\end{itemize}

We will prove that $r|p^n-1$, provide an explicit formula for $a$ and show that $\beta:= n \log_p{a}/(n')^2 \leq 1 $.

We will first show that  $r=\frac{(p^n-1)(k-(-1)^n)}{(k^{n}-k)(k^n-(-1)^n)}$
is an integer dividing $p^n-1$.
We can notice that $(k^{n}-k)|p^n-1$. Indeed we have :
\begin{align*}
(p^n-1)&=((k^n-k)+(-1)^{n+1})^n-1\\
&=\sum_{i=1}^{n}\binom{n}{i}(k^n-k)^i(-1)^{(n+1)(n-i)}+ (-1)^{n(n+1)}-1\\
&=(k^n-k)\sum_{i=0}^{n-1}\binom{n}{i}(k^n-k)^i(-1)^{(n+1)(n-i)}
\end{align*}

Similarly, $(k^{n}-(-1)^n)|p^n-1$ since
\begin{align*}
         (p^n-1)&=((k^n-(-1)^{n})-k)^n-1\\
&=\sum_{i=1}^{n}\binom{n}{i}(k^n-(-1)^n)^i(-k)^{n-i}+(-k)^n-1\\
&=\left(k^n-(-1)^n\right)\left(\sum_{i=0}^{n-1}\binom{n}{i}(k^n-(-1)^n)^{i}(-k)^{n-i}+(-1)^n\right)
\end{align*}

Therefore $(k^{n}-k)(k^{n}-(-1)^n)/\gcd(k^{n}-k,k^n-(-1)^n)$ is an integer dividing $p^n-1$.
Moreover, 
\begin{align*}
	\gcd(k^{n}-k,k^n-(-1)^n)&=\gcd((k^n-(-1)^n)-(k^{n}-k),k^n-(-1)^n)\\&=\gcd(k-(-1)^n,k^n-(-1)^n)\\
	&=k-(-1)^n \text{ since } k-(-1)^n|k^n-(-1)^n.
\end{align*}

Hence $p^n-1= r .\dfrac{(k^{n}-k)(k^{n}-(-1)^n)}{\gcd(k^{n}-k,k^n-(-1)^n)}$ and thus $r|p^n-1$

Since the value of $r$ depends on the parity of n, we  distinguish two cases for the remaining of the proof.

\paragraph{If $n$ is even}
We now show that $a:=(p^{n'} \mod r) = \frac{p^{n-1}+1}{k^n-k}$
Indeed, $\frac{p^{n-1}+1}{k^n-k} = p^{n-1}-\frac{p^n-1}{k^n-k}=
p^{n'}-r \frac{k^n-(-1)^n}{k-(-1)^n}$. Moreover, $\frac{p^{n-1}+1}{k^n-k}\simeq k^{n(n-1)-n} = k^{n^2-2n}$
 while $r\approx k^{n^2+1-2n} $ for $k^n\gg 1$, so $\frac{p^{n-1}+1}{k^n-k}\simeq r/k$ 
 	and thus for $k$ big enough, we have $\frac{p^{n-1}+1}{k^n-k}<r$.

Furthermore, $p^{n'}-a=\frac{p^n-1}{k^n-k} |p^n-1 $ so $L$ splits over $\mathbb{F}_{p^n}$ and it has  $p^{n'}-a +1= \frac{p^n-1}{k^n-k}+1 \simeq k^{n(n-1)}$ roots. This is close to $p^{n'}$ roots if $k^n\gg 1$.

Finally, $\frac{a^n}{p^{(n-1)^2}}\simeq \frac{(k^{n^2-2n})^n}{k^{n(n-1)^2}} = (k^n)^{(n^2-2n-n^2+2n+1)} =k^{-n}<1 $ when $k^n\gg 1$ so $a^n<p^{(n-1)^2}$, thus $$\beta = n \log_p{a}/(n-1)^2 = (\log_p{a^n})/(\log_p{p^{(n-1)^2}})\leq 1.$$

\paragraph{If $n$ is odd}
We now show that $a=\frac{p^{n-1}k+1}{k^n+1}$.

Indeed $\frac{p^{n-1}k+1}{k^n+1}=p^{n-1}-\frac{p^n-1}{k^n+1}=p^{n'}-r \frac{k^n-k}{k-(-1)^n}$ and we have

\begin{equation*}
    \begin{split}
r&=\frac{(p^n-1)(k+1)}{(k^{n}-k)(k^n+1)}\\ 
&= \frac{p^{n-1}}{k^n+1} (p(k+1))\frac{1}{k^n(1-k^{1-n})} -\frac{1}{(k^{n}-k)(k^n+1)}\\
&=\frac{p^{n-1}}{k^n+1} (k^{n+1}+k^{n}+o(k^3))k^{-n}(1+k^{1-n}+o(k^{1-n})) +o(1) \text{ for } k^n\gg 1\\
&=\frac{p^{n-1}}{k^n+1} (k+1+o(k^{3-n}))(1+k^{1-n}+o(k^{1-n})) +o(1)\\
&=\frac{p^{n-1}}{k^n+1} (k+1+o(k^{3-n})) +o(1)\\
&=\frac{p^{n-1}k+1}{k^n+1} -\frac{1}{k^n+1}+\frac{p^{n-1}}{k^n+1} (1+o(k^{3-n})) +o(1)\\
& =\frac{p^{n-1}k+1}{k^n+1}+\frac{p^{n-1}}{k^n+1} (1+o(k^{3-n})) +o(1)\\
& = \frac{p^{n-1}k+1}{k^n+1}+ \frac{p^{n-1}}{k^n+1} (1+o(1))        \text{ since } n\geq 3, \\
& > \frac{p^{n-1}k+1}{k^n+1}
 \end{split}
\end{equation*}
Moreover, $p^{n'}-a=\frac{p^n-1}{k^n+1}|p^n-1$ so $L$ splits over $\mathbb{F}_{p^n}$ and it has 
$p^{n'}-a +1= \frac{p^n-1}{k^n+1}+1 \approx k^{n(n-1)} \approx p^{n'}$ roots if $k^n\gg1$.

Furthermore,
\begin{equation*}
    \begin{split}
a^n&=\left(\frac{p^{n-1}k+1}{k^n+1}\right)^n=\left(\frac{p^{n-1}k}{k^n+1}+o(1)\right)^n\\
    &=\frac{p^{(n-1)^2+n-1}k^n}{(k^n+1)^n}+o\left(\frac{p^{n(n-1)}k^n}{k^{n^2}}\right)\\
&=p^{(n-1)^2}\frac{p^{n-1}k^n}{(k^n+1)^n}+o(k^{n(n-1)^2}) \\&=p^{(n-1)^2}\frac{(k^{n(n-1)}-(n-1)k^{1+n(n-2)}+o(nk^{1+n(n-2)})) k^n}{(1+1/k^n)^n} k^{-n^2}\\
& \quad \quad \quad \quad +o(k^{n(n-1)^2})\\
&=p^{(n-1)^2}(1-(n-1)k^{1-n}+o(nk^{1-n})) (1-k^{-n}+o(k^{-n}))+o(k^{n(n-1)^2})\\
&=p^{(n-1)^2}(1-(n-1)k^{1-n}+o(nk^{1-n}))+o(k^{n(n-1)^2})\\
&<p^{(n-1)^2}
 \end{split}
\end{equation*} 

Thus $\beta := n \log_p{a}/(n-1)^2 = (\log_p{a^n})/(\log_p{p^{(n-1)^2}})\leq 1 $.

\end{document}